\documentclass[a4paper,11pt]{article}
\usepackage{jcappub} 
\usepackage{lineno}
\usepackage{soul}

\newcommand{\hlcolor}[2][yellow]{{#2}}
\newcommand{\hlc}[2][yellow]{{#2}}

\title{Average energy of the X-ray spectrum as a model-independent proxy for the mass of galaxy clusters}

\author[a]{Aleksei Kruglov,}
\author[b, c, a]{Ildar Khabibullin,}
\author[a,d]{Natalya Lyskova,}
\author[b,c]{Klaus Dolag,}
\author[e,f]{Veronica Biffi}

\affiliation[a]{Space Research Institute (IKI),\\
Profsoyuznaya 84/32, Moscow 117997, Russia}
\affiliation[b]{Universit\"ats-Sternwarte, Fakult\"at f\"ur Physik, Ludwig-Maximilians-Universit\"at M\"unchen,\\
Scheinerstr.~1, D-81679 M\"unchen, Germany}
\affiliation[c]{Max Planck Institute for Astrophysics,\\ 
Karl-Schwarzschild-Str. 1, D-85741 Garching, Germany}
\affiliation[d]{Astro Space Centre, P.N. Lebedev Physical Institute,\\
Profsoyuznaya 84/32, Moscow 117997, Russia}
\affiliation[e]{INAF, Osservatorio Astronomico di Trieste,\\ 
via Tiepolo 11, I-34131, Trieste, Italy}
\affiliation[f]{IFPU – Institute for Fundamental Physics of the Universe,\\
Via Beirut 2, I-34014 Trieste, Italy}

\emailAdd{kruglov@cosmos.ru, ildar@mpa-garching.mpg.de, lyskova@cosmos.ru}

\abstract{Temperature of the hot gas in galaxy clusters is known to be a reliable proxy for their total gravitating mass, allowing one to use spectroscopic X-ray observations for halo mass function measurements. Data of shallow wide area surveys, however, often precludes direct fitting of the X-ray spectra, given possible biases arising due to unresolved (multi-temperature) inner structure of the intracluster medium (ICM), projection effects and necessity of certain model assumptions to be made to allow for robust spectral fitting.
We consider using a simple observable value - the average energy of the observed cluster X-ray spectrum - as a model-independent proxy for the ICM temperature, and consequently cluster's mass. We calibrate relation of this proxy to the cluster parameters using mock observations for a sample of 84 massive galaxy clusters extracted from the \textit{Magneticum} cosmological hydro simulations. We consider observational parameters corresponding to the all-sky survey observations by \textit{SRG/eROSITA}. Taking into account contributions of various background and foreground signals, average energy of the simulated X-ray spectra in the $0.4-7.0$ keV band is shown to be a stable indicator of the ICM temperature with $\sim10\%$ scatter and cluster's mass $M_{500}$ with a $\sim 20\%$ scatter. A database containing simulated X-ray images and their spectra (subtracted in several concentric rings) is publicly available.}

\begin{document}
\maketitle
\flushbottom


\section{Introduction}


The number density of massive (with virial mass $\gtrsim 10^{14}$M$_\odot$) galaxy clusters and their spatial distribution reflect the cosmological evolution of the Universe, governed in particular by its dark matter and dark energy content, as well as the power spectrum of the initial density perturbations. Thanks to that, measurements of the halo mass function (i.e. number density as a function of mass) for galaxy clusters have been successively used to infer values of corresponding parameters of the $\Lambda$ Cold Dark Matter ($\Lambda$CDM) cosmological model, in particular relative matter $\Omega_{\rm M}$ and dark energy $\Omega_{\rm \Lambda}$ density or amplitude of linear density fluctuations $\sigma_8$ \cite[][for a recent review]{2012ARA&A..50..353K}. 

Masses of galaxy clusters\footnote{As a galaxy cluster halo doesn't have a distinct boundary, the definition of cluster mass needs to be clarified. To avoid ambiguity, throughout this paper this quantity will be considered as $M_{500}$, i.e. the mass enclosed inside the sphere of radii $R_{500}$ (within which the mean density is equal to 500 times the critical density of the Universe at the epoch of observation). This choice of overdensity is a compromise between theoretical requirements (driven by numerical resolution in simulations, accounting for non-gravitational effects, hydrostatic equilibrium assumptions etc.) and observational uncertainties, which inevitably arise for smaller values of overdensity.} can be determined directly only through gravitational lensing measurements, which, however, are accompanied by systematic effects such as the presence of large-scale structures along the line-of-sight, assumed underlying dark matter distribution, and shear from massive substructures \citep[e.g.][]{2010A&A...514A..93M, 2003MNRAS.339.1155H}.  On the other hand, the cluster mass can be determined implicitly from properties of its intracluster medium (ICM), which can be inferred from observations of its thermal X-ray emission \citep[e.g.][]{1986RvMP...58....1S, 2010A&ARv..18..127B} or Sunyaev-Zeldovich distortion of the Cosmic Microwave Background radiation \citep{1972CoASP...4..173S} in their direction. 

Several quantities derivable from X-ray and microwave observations were put forward as proxies for cluster mass estimation, among them the total X-ray luminosity \citep{1986MNRAS.222..323K}, the spectral temperature of the thermal X-ray emission \citep{1998ApJ...504...27M, 1999MNRAS.305..631A, 2006ApJ...640..691V},  the mass estimate of the X-ray emitting gas \citep{2003ApJ...590...15V}, or the so-called $Y_X$ parameter \citep[the product of the ICM mass and its temperature,][]{2006ApJ...650..128K}.  Accurate measurements of these quantities for large samples of galaxy clusters are, however, challenging, especially in the regime of low mass clusters and shallow observations delivered by modern wide area surveys.

Moreover, given the multi-temperature inner structure of the ICM, its average temperature could be defined in several ways. For instance, one can use data of cosmological hydrodynamical simulations that carry information about density, temperature and volumetric density of every ICM mass element to derive volume or mass-averaged temperature of the hot gas. Given that the luminosity of the thermal X-ray emission scales with the square of the gas density, one can also consider the emission measure-weighted temperature. Observationally, the X-ray temperature is inferred by fitting the spectrum of X-ray emission with single- or multi-temperature collisional ionization equilibrium (CIE) plasma emission models \citep[e.g. APEC,][]{2012ApJ...756..128F} with free or fixed abundance of the heavy elements. However, in many cases, the statistical errors in the observed spectrum are too large to detect the presence of multiple emission components. Additional bias may be introduced by projection effects, i.e. accounting for photons from structures along the line-of-sight \citep{2023A&A...675A.150Z}. X-ray telescope optics, namely, the energy-dependent effective area, is another source of temperature uncertainties \citep{2015A&A...575A..30S, 2024A&A...688A.107M}. It has been shown that the temperature derived from X-ray observations tends to be different from any of the defined temperatures derived from simulations, generally underestimating them \citep{2001ApJ...546..100M,2004MNRAS.354...10M}.

Compared to the cluster temperature, the X-ray luminosity (bolometric or in a certain energy range, for example $0.5-2.0$ keV) is a more straightforward mass proxy. However, for a fixed cluster mass, this quantity demonstrates the relatively large scatter ($\sim$ 40 \%) since it is very sensitive to the exact properties of the gas near the cluster core. With sufficient angular resolution, it is possible to exclude emission from the cluster nuclei, thereby reducing the scatter \citep[e.g.][]{1998ApJ...504...27M, 2009A&A...498..361P, 2019ApJ...871...50B, 2020OJAp....3E..12E}.
Similarly to temperature, luminosity derived from simulations and observations may differ by a few percent, and this discrepancy inevitably follows the choice of the observed region; additionally, the fitting procedure required for luminosity calculations will not capture the full expected luminosity, as it requires sufficient statistics to constrain the metallicity parameter \citep{2023A&A...675A.150Z}.

In this work, we propose using an average energy of the X-ray spectrum (i.e. the value obtained by averaging nominal energies of each detector channel with weights equal to the observed count rate; see Sec.~\ref{sec:average_energy}) of galaxy clusters as a temperature and mass proxy. The major advantage of the presented approach is that it does not require any spectral fitting and is less prone to possible biases due to the multiphase structure of the gas, assumptions about cluster metallicity, and impact of masking of certain bright and relatively cool regions of ICM. In other words, the average energy works reasonably well as a mass/temperature proxy even in a case of low count statistics (at the level of a few hundred total counts which are enough to identify X-ray source as an extended one but not enough for robust spectral analysis) and provides a scatter that is comparable to mass measurements derived from temperature or luminosity. We take advantage of mock observations for a sample of galaxy clusters from the \textit{Magneticum}\footnote{\href{http://www.magneticum.org/}{\texttt{http://www.magneticum.org}}} suite of state-of-the-art cosmological hydrodynamical simulations modeled for the parameters of the all-sky survey by \textit{SRG/eROSITA} telescope \citep{2021A&A...647A...1P} of the \textit{Spectrum-RG} observatory \citep{2021A&A...656A.132S}. Via analysis of the predicted X-ray spectra, we calibrate the relationship between the average energy of galaxy cluster spectra and cluster temperatures and masses, and compare statistical properties of this relation to other commonly used proxies.

The paper is structured as follows: in Section~\ref{sec:sample} we briefly describe the sample of clusters used for our study and compare the $T_{500}-M_{500}$ and $L_{\rm x500}-M_{500}$ relations (where $T_{500}$ is the mass-weighted temperature and $L_{\rm x500}$ is the bolometric X-ray luminosity, both inside the sphere of $R_{500}$) for them with the scaling relations derived for observed clusters.
In Section~\ref{sec:xray}, we perform modeling of realistic X-ray spectra, fit a single-temperature model and derive a relation between the average energy of observed cluster spectra and the cluster temperature. In Section~\ref{sec:discussion}, we compare luminosity, temperature and average energy as mass proxies. Finally, in Section~\ref{sec:conclusions}, we summarize our conclusions.

\section{Cluster sample}\label{sec:sample}

\subsection{\textit{Magneticum} Simulations}
For our study we have used high-resolution numerical cosmological hydrodynamic simulations \textit{Magneticum} \citep{2014MNRAS.442.2304H,2016MNRAS.463.1797D}. These simulations were performed with the TreePM/SPH code \texttt{GADGET-3} \cite[][]{2005MNRAS.364.1105S, 2016MNRAS.455.2110B, 2023MNRAS.526..616G}.
They take into account many complex non-gravitational physical processes (cooling, merging of galaxies and galaxy clusters, shock waves, and detailed galaxy formation physics) which determine the evolution of large-scale systems and affect their observational properties. More details of the physics provided in the simulation can be found in Section 2 of \cite{2022A&A...661A..17B} and references therein.

For this research we have chosen \textit{Box2/hr} simulation box. \hlc[yellow]{A cluster sample drawn from this particular simulation box was chosen for two reasons: (a) it encompasses a sufficiently large volume to include a statistically significant number of massive galaxy clusters (in particular, $>10^{14} M_{\odot}$); (b) it provides a reliable compromise between simulated size and resolution required for adequate modelling of the impact of galaxy evolution and feedback processes on ICM properties.} Its comoving volume is equal to $(352 \text{Mpc}/h)^3$ ($h=H_0/{\rm (100~km s^{-1} Mpc^{-1})}$ is the Hubble constant), containing $2\cdot1584^3$ mass resolution elements (particles). The masses of dark matter and gas particles are equal to $m_{\rm DM} = 6.9\cdot10^8 M_{\odot}/h$ and $m_{\rm gas} = 1.4\cdot10^8 M_{\odot}/h$, respectively, and the following cosmological parameters are adopted: the total matter density $\Omega_{\rm M}=0.272$ ($16.8\%$ baryons), the cosmological constant $\Omega_{\Lambda}=0.728$, the Hubble constant $H_0=70.4$ km/s/Mpc (i.e. $h=0.704$), the index of the primordial power spectrum $n=0.963$, the overall normalisation of the power spectrum $\sigma_8=0.809$ \citep{2011ApJS..192...18K}.

\begin{figure}[h]
    \centering
    \includegraphics[width=0.7\textwidth]{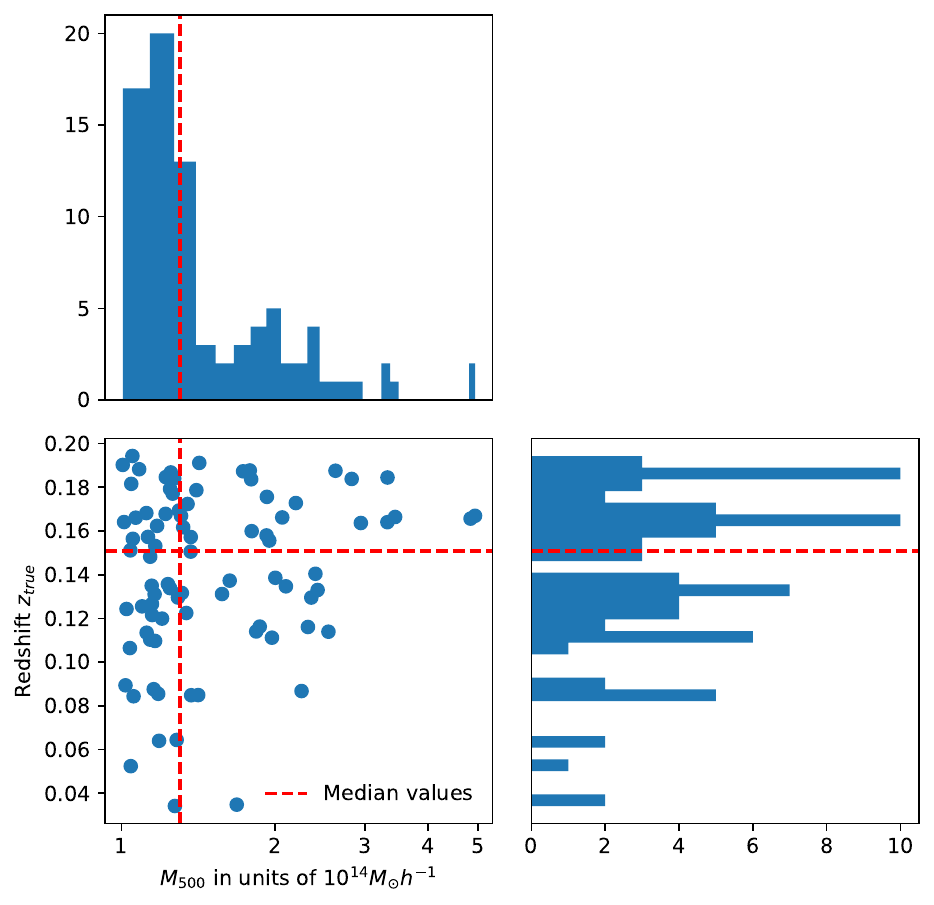}
    \caption{Histograms of masses $M_{500}$ and redshifts $z_{\rm true}$ (see text for clarification) for 84 clusters in the sample. Red dashed line shows median values: $M_{500}^{med} = 1.30 \cdot 10^{14} M_{\odot}/h$ and $z_{\rm true}^{med} = 0.15$.
    }
    \label{fig:histograms_84}
\end{figure}

By means of the \texttt{PHOX} software package \citep{2012MNRAS.420.3545B,2013MNRAS.428.1395B,2023A&A...669A..34V}, a  $30 \times 30$ degrees lightcone was {constructed} based on the \textit{Magneticum} \textit{Box2/hr} simulation. {To produce such mock X-ray data, first {\it unit1} of \texttt{PHOX} was used to create photons emitted by the gas mass elements (including ICM), assuming optically thin thermal emission and taking into account the line emission of the various individual metal species tracked within the simulation according to the metal production yields. We neglect the predicted X-ray emission from the AGN component \citep[][]{2018MNRAS.481.2213B} within the simulation although it is in principle available, as well as the contribution of the X-ray binaries in cluster galaxies \citep{2023A&A...669A..34V}. Then {\it unit2} of \texttt{PHOX} was used to create the photon list for a lightcone-like geometry, using simulation outputs at the following redshifts $z=\{0.033, 0.066, 0.101, 0.137, 0.174\}$ to compose individual slices. The corresponding redshift within each slice is properly computed from the offset with respect to the center of the slice, and the photon energies are corrected for the redshift and the peculiar velocity of every resolution element within the lightcone. \hlc[yellow]{Foreground absorption by Galactic neutral hydrogen is included with a \texttt{wabs} absorption model }\cite{1983ApJ...270..119M}\hlc[yellow]{ with column density $N_H=10^{20} \ \rm cm^{-2}$. The photons produced by \texttt{PHOX} assume the following observation setup: observation time of $10 \ \rm ks$ and an energy-independent effective area of $1000 \ \rm cm^2$.} In addition, the corresponding metadata for the galaxy clusters within the lightcone have been created, including the redshift $z_{\rm true}$ (e.g. the purely cosmological) and $z_{\rm observed}$ (e.g. including the peculiar velocity of the cluster). The data is publicly available and can be downloaded from the data section of the {\it Magneticum} project webpage. Based on these halo catalogues provided for each slice of the lightcone \hlc[yellow]{(which include 19961 haloes in total)}, we have selected all 84 massive clusters with \hlc[yellow]{masses $M_{500}$ above} $10^{14} M_{\odot}/h$} as shown in Figure~\ref{fig:histograms_84}.

Several recent studies have taken advantage of a similar dataset from the Magneticum simulations to draw predictions regarding X-ray emission from galaxy clusters and groups. In particular, \cite{2023A&A...675A.150Z} investigated the bias in temperature and luminosity estimates arising due to single-temperature spectral approximation and projection effects, having shown that they might indeed be significant. \cite{2022A&A...663L...6A} have studied the properties of baryons in halos with $M_{\rm vir} > 10^{13} M_{\odot}/h$ far beyond the virialised regions and have shown that, despite different ways of convergence which depend on their masses, these haloes have common baryon, gas, stellar and metal distributions as they approach the mean location of the accretion shock. \cite{2024A&A...689A...7M} have confirmed the high completeness and purity of the detection of massive galaxy clusters ($M_{500} > 10^{14} M_{\odot}$) detection with \textit{SRG/eROSITA}. \cite{2023A&A...670A..33S} have investigated hydrostatic equilibrium in clusters of galaxies and recovered temperature profiles with an accuracy of the order of $~1\%$, providing predictions for the hydrostatic mass bias.
\cite{2023A&A...671A..57S} have concluded that there is good agreement in the offsets between X-ray and optical centers in clusters of galaxies measured in \textit{SRG/eROSITA} data and the predictions from the Magneticum simulations.

The basic properties of the simulated clusters that are relevant to our study, such as
their positions inside the simulation box, redshifts $z_{\rm true}$, radii $R_{500}$, masses $M_{500}$, mass-weighted temperatures $T_{500}$ and bolometric X-ray luminosities $L_{\rm x500}$ are summarized in a catalogue that is made publicly available and to which we will refer throughout this paper.

\subsection{Scaling relations}\label{sec:sc_rel}

As one of the key goals of the current study is to obtain relations between observable spectral properties of the X-ray emission and parameters of the clusters such as ICM temperature and total mass, we first consider the intrinsic correlations between these quantities and their consistency with the currently available observations.  

In order to do so, we show how cluster parameters from the catalogue, namely $T_{500}$ and $L_{\rm x500}$, correlate with the cluster mass $M_{500}$. We adopt the scaling relations for these quantities from \cite{2009ApJ...692.1033V} derived for high-quality \textit{Chandra} observations of clusters with $0.02 < z < 0.9$ and $M_{500} \gtrsim (1-2)\times10^{14} h^{-1} M_{\odot}$ in the form presented in \cite{2015MNRAS.450.1984C}. In these expressions, $h = H_0/100 \ \textrm{km/s/Mpc} \equiv 0.704$ and $E(z) = \sqrt{\Omega_{\rm M} (1+z)^3 + \Omega_{\Lambda}}$ where $\Omega_m = 0.272, \Omega_{\Lambda} = 0.728$ \hlc[yellow]{(we assume $\Omega_{\rm k}=0$, i.e. flat cosmology)} and $z$ is the cluster redshift $z_{\rm true}$.

According to \cite{2009ApJ...692.1033V}, the scaling between mass $M_{500}$ and temperature is given by the following expression

\begin{equation} \label{eq:T_X}
    T_X = 5 \ \text{keV} \Bigg( \frac{M_{500}}{M_0} \Bigg)^{0.65} E(z)^{0.65}
\end{equation}
where $M_0 = (2.95 \pm 0.10) \ 10^{14} \ h^{-1} M_{\odot}$, $T_X$ is the gas temperature derived from a single temperature fit to the total cluster spectrum integrated within $(0.15-1) \cdot R_{500}$ radial range. The scatter of $T_X$-based mass estimations for this relation was adopted equal to $20\%$ based on simulations from \cite{2006ApJ...650..128K}.

For the X-ray luminosity, the scaling relation from \cite{2009ApJ...692.1033V} reads as

\begin{equation} \label{eq:L_X}
    L_{\rm X, 0.5-2} = 1.35 \cdot 10^{44} \ \text{erg s}^{-1} \Bigg( \frac{h}{0.72} \Bigg)^{-0.39} \Bigg( \frac{M_{500}}{M_0} \Bigg)^{1.61} E(z)^{1.85}
\end{equation}

where $L_{\rm X,0.5-2}$ is the rest frame cluster luminosity in the $0.5-2$ keV band. For a fixed mass, the scatter in $L_X$ reaches $\approx 48\%$.

\hlcolor[yellow]{Besides the statistical and intrinsic scatters in the observationally-derived scaling relations, an important additional complication comes from systematic biases inherent to the particular methods (and assumptions on which they are based) for inferring corresponding values. A very well known example is the so-called hydrostatic mass bias arising due to the assumption of hydrostatic equilibrium between the gas thermal pressure gradient and the gravitational pull of the halo} (e.g. \cite{2016ApJ...827..112B}). \hlcolor[yellow]{Although such systematic effects can, in principle, be pinned down by cross-calibration of different measurement methods, the resulting accuracy is still limited, with the accuracy levels better than 10\% being very challenging to achieve. In order to illustrate the effect of such possible miscalibrations, one might introduce a close-to-unit correction factor $\eta$ for the true halo mass compared to the one derived in observations (and used in derived scaling relations). It is the former mass that would provide proper cosmological halo mass inference and comparison with the numerical simulations, but for the simpler task of approximate mass determination (given limited and noisy observables), the deviation of this factor from unity plays a relatively small role.}

\begin{figure*}[h]
  \includegraphics[width=\textwidth]{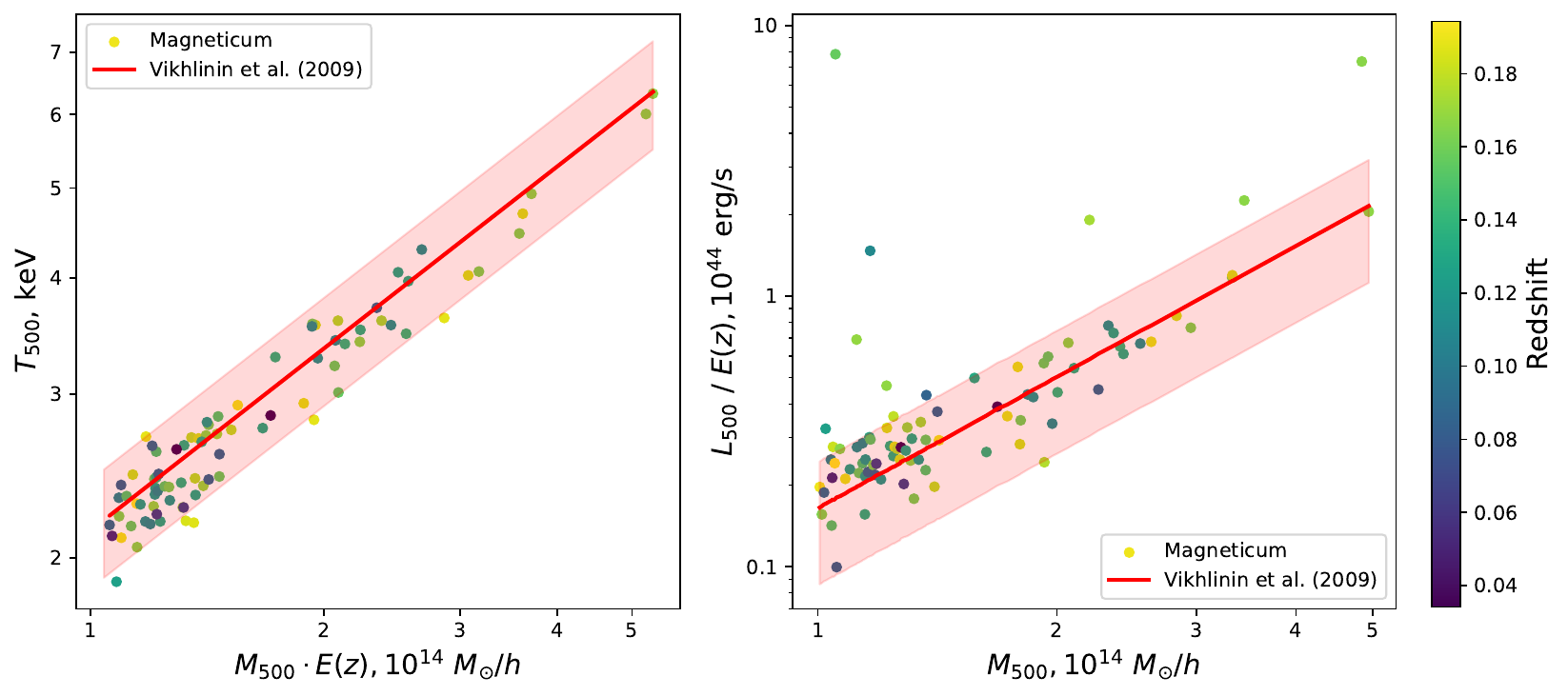}
  \caption{Scaling relations for 84 clusters from our sample. Points indicate the clusters from the \textit{Magneticum} catalogue. Red lines show expressions (\ref{eq:T_X}) and (\ref{eq:L_X}), shaded areas show scatter from \citep{2009ApJ...692.1033V}. \hlcolor[yellow]{The masses $M_{500}$ in scaling relations were corrected by the factor $\eta=0.8$ (see text for explanation).} Redshifts of clusters are indicated by colorbar.}
  \label{fig:scaling_relations}
\end{figure*}

In Fig.~\ref{fig:scaling_relations} we illustrate the scaling relations from \cite{2009ApJ...692.1033V} \hlcolor[yellow]{assuming $\eta = 0.8$ (e.g. }\cite{2016ApJ...827..112B}), \hlcolor[yellow]{ or, in other words, that the mass estimates from scaling relations} \ref{eq:T_X} and \ref{eq:L_X} \hlcolor[yellow]{should be divided by $\eta$ to be consistent with the total halo masses from simulations. The same figure shows the properties of our sample,} \hlc[yellow]{indicating an even tighter correlation, since almost all clusters from our sample are within the scattering envelope}. It is necessary to recall that $T_{500}$ from the catalogue is a mass-weighted temperature within $R_{500}$, while the observationally-derived $T_X$ is a single temperature fit to the total cluster spectrum. \hlcolor[yellow]{Therefore, perfect agreement between simulated and observed scalings is not expected.} Also, $L_{500}$ is the bolometric X-ray luminosity within $R_{500}$ and therefore it does not strictly correspond to $L_{\rm X, 0.5-2}$. To compare the latter quantity with $L_{500}$, we have done simple rescaling by calculating $L_{\rm X, 0.5-2}$ as a fraction of the full luminosity for simulated spectra of \texttt{APEC} model \citep{2012ApJ...756..128F} for the cluster with given temperature $T_X$ and redshift $z_{\rm true}$ from the catalogue. The right panel of Fig.~\ref{fig:scaling_relations} shows acceptable compliance of the $L_{\rm x500}-M$ relation for our sample of clusters, although there are some significant outliers.

\hlc[yellow]{Recently, the scaling relations for haloes in Magneticum simulations have been extensively discussed, e.g. in }\cite{2022MNRAS.517.5303L} (temperature-mass), \cite{2024arXiv241117120P} (luminosity-mass), \hlc[yellow]{and they have also found good agreement with observations.}

\section{X-ray spectral analysis}\label{sec:xray}

\subsection{Generating clusters images and spectra}\label{sec:im_gen}

For each cluster, we generate its image with a resolution equal to $4$ arcseconds/pixel. We take the position of the centroid of the image (which is calculated by weighing the values of surface brightness in each pixel) as a reference point. This is done instead of pointing at the gravitational potential minimum from the catalogue of clusters, which corresponds to the position of the main subhalo identified by the \texttt{SubFind} algorithm \citep{2001MNRAS.328..726S, 2009MNRAS.399..497D}. Despite the fact that $R_{500}$ (but also $M_{500}$ and $T_{500}$) from the simulations refer to the overdensity calculated considering that the center of the cluster is located in the minimum of the potential well, the possible bias is expected to be small \citep{2023A&A...671A..57S}.

We collect all photons within a cylinder with radius equal to $R_{500}$ from the catalogue divided by the angular diameter distance to the cluster obtained with redshift $z_{\rm true}$ from the catalogue as well. {The photons are collected only from the same slice of the lightcone to which the cluster belongs (i.e., the photons from other slices are not taken into account). However, the line-of-sight thickness of the slices, $\delta z\approx0.033$, corresponds to $\gtrsim100$ Mpc, ensuring that the correlated large-scale structures surrounding these massive clusters are included in the extracted photon lists because of the projection effect as well.} For simplicity, we consider the whole sample of model clusters, although it includes objects at different dynamical and merger states.

The cluster images may contain some structures that obviously stand out from the bulk ICM, for example, gas-rich galaxies. The excess of soft X-ray emission from these small "clumps" of cooler gas could have a viable contribution to the total spectrum of the cluster and, therefore, can strongly shift the results of spectral analysis, leading to temperature underestimation \citep{2004MNRAS.354...10M}. To mitigate possible undesirable effects, we do not take into account areas that contain $5\%$ of the brightest pixels. The loss of the total photon count rate due to this filtering is $25\%$ on average, although for some images it may reach $\sim40\%$. However, this procedure leads to the extraction of a minimally contaminated spectrum, which, as will be shown below, is quite well described by the single-temperature approximation \citep[a similar filtering was considered and quantified in][based on the cosmological simulations]{2013MNRAS.428.3274Z}. Examples of original and filtered cluster images are shown in Appendix~\ref{sec:database}.

The energies of the photons collected from the circle of $R_{500}$ centered at the centroid of the X-ray image are multiplied by $(1+z_{\rm true})$, i.e. we consider that all clusters are observed at $z=0$ to mitigate possible effects of redshift onto future analysis. The X-ray flux density spectra for each cluster are produced and then passed as input model spectra to the \texttt{XSPEC} software \citep{1996ASPC..101...17A}. We present\footnote{\href{https://github.com/pi4imu/RRCS_DB}{\texttt{https://github.com/pi4imu/RRCS\_DB}}} Radially-Resolved Clusters' Spectra database (RRCS\_DB) implemented as a single \texttt{XSPEC} table model, which includes model spectra for all studied clusters along with their images (both original and filtered). The description of this database is given in Appendix~\ref{sec:database}.

\subsection{Mock observations and fitting overview}\label{sec:overview}

We simulate galaxy cluster observations by \textit{SRG/eROSITA} telescope, considering them to be performed by the telescope module TM1 only, but with a significantly long exposure equal to $10$ ks. This approximately corresponds to the effective exposure per telescope module of \textit{SRG/eROSITA} during four consecutive all-sky surveys \citep{2023MNRAS.525..898L, 2024A&A...682A..34M}.

For each cluster, its synthetically generated model spectrum in the $0.1-12$ keV band is convolved with the \textit{SRG/eROSITA} TM1 RMF (\textit{Redistribution Matrix File}) and ARF (\textit{Auxiliary Response File}) with the addition of statistical fluctuations described by Poisson errors using the \texttt{fakeit} command provided by \texttt{XSPEC} software. The resulting count spectrum is limited to the $0.4-7.0$ keV range: the upper border is motivated by the presence of the bright line at 6.7 keV (helium-like iron Fe XXV) and the lower border is limited by the prevalence of cosmic X-ray background (which will be added to the cluster spectra later) at smaller energies. The mock observation spectra are approximated by the single-temperature \texttt{phabs*apec} model with the hydrogen column density of $N_H=10^{20} \ \rm cm^{-2}$ (since initial photons were simulated with this value of absorption) and the fixed abundance $Z=0.3 \cdot Z_{\odot}$ (this choice is motivated by the results of numerical simulations \cite[e.g.][]{2017MNRAS.468..531B}; also, our tests have shown that fixing it does not lead to any significant bias in the result). We set $z=0$ for each cluster during the fitting procedure because the energies of photons have already been redshifted; temperature and normalization parameters are left free. C-statistics is applied and the weights are taken as \texttt{standard}. Examples of X-ray cluster spectra along with the best-fit model (with and without adding an X-ray background, which is described in the following) are shown in Appendix~\ref{sec:database}. We denote the temperature from the best-fit model as $T_{\rm spec}$.

This process is repeated 50 times for each cluster, and we further consider $T_{\rm spec}$ as the mean value of the spectral temperature distribution. \hlc[yellow]{In order to properly estimate the mean error on temperature, }one uncorrected standard deviation \hlc[yellow]{of the spectral temperature distribution} is taken as the uncertainty \hlc[yellow]{of $T_{\rm spec}$}, albeit these distributions are not necessarily Gaussian even in case of sufficiently many realizations, which is demonstrated in Appendix~\ref{sec:howmuch}. Figure~\ref{fig:T_vs_T500} shows a comparison between $T_{\rm spec}$ from the single-temperature approximation and the mass-weighted $T_{500}$ temperature taken from the catalogue for our sample of clusters. These two values are in good agreement, which can be explained by the previously described filtering of images: relatively small bright sources that could be the cause of the multi-temperature structure of the cluster were partially or fully excised. Also, there is no significant trend that links the cluster's redshift and its temperature, whereas errors on temperature grow almost linearly with redshift.

\begin{figure}[h!]
  \centering
  \includegraphics[width=\textwidth]{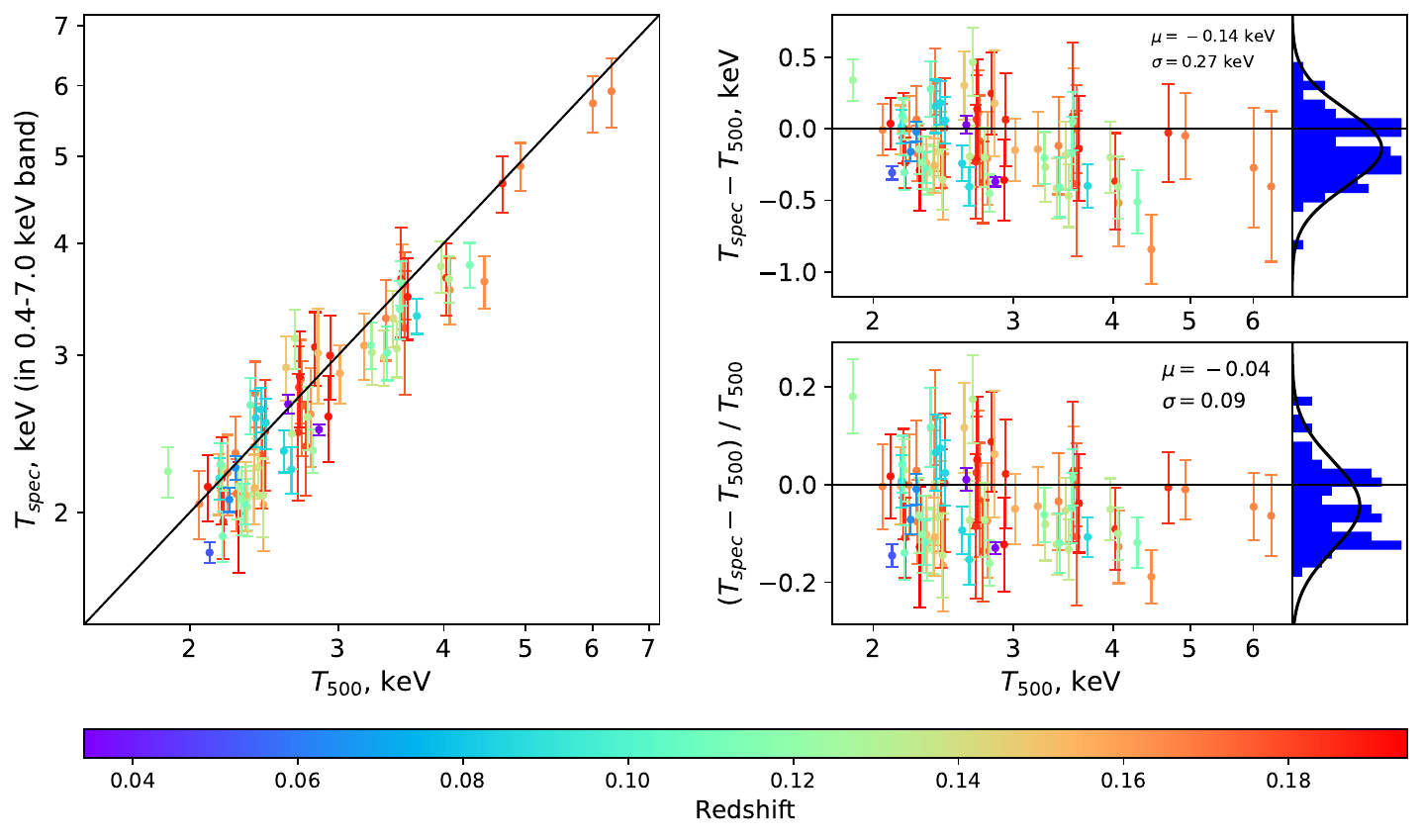}
  \caption{Comparison of the mass-weighted temperature $T_{500}$ from the simulations with $T_{\rm spec}$ from spectral fitting for 84 clusters from \textit{Magneticum} simulations. The upper panel compares temperatures directly, whereas central panel compares absolute difference and lower panel compares relative difference between them. The histograms on the right indicate mean values $\mu$ and scatter $\sigma$ for this comparisons. Redshifts are indicated by colorbar. Note the log-scale for the horizontal axis.}
  \label{fig:T_vs_T500}
\end{figure}

Simultaneously with spectral fitting, for each cluster and for each realisation we calculate the X-ray luminosity in the $0.5-2.0$ keV band. Again, the mean value of its distribution is denoted as $L_{\rm spec}$ and one standard deviation is considered as an uncertainty. The results of this performance are described in Section~\ref{sec:discussion}. From mock observed spectra we also derive the value of the average energy, which is described in the next section.

\subsection{Average energy}\label{sec:average_energy}

We suggest yet another indicator of a cluster gas temperature - the average energy of counts in the observed X-ray spectra. In this work, if not stated otherwise, we consider this quantity in the same energy band that was used for spectral fitting, i.e. in $0.4-7.0$ keV. Based on our mock observations, we calculate the average energy $E_{\rm av}$ of the observed spectra following \cite{2006ApJ...640..710VF}:

\begin{equation} \label{eq:aven}
    E_{\rm av} = \frac{\sum E_i s_i}{\sum s_i},
\end{equation}
where $s_i$ is the observed count rate in the channel $i$ and $E_i$ is the nominal energy corresponding to this channel. The count rates $s_i$ depend on the temperature, interstellar absorption, and detector sensitivity as a function of energy. Therefore, the average energies of an observed spectrum calculated for different detectors/telescopes are not necessarily the same. \cite{2006ApJ...640..710VF} used this value as an indicator of temperature only in line-dominated spectra. Here, we  extend the applicability of this quantity to realistic X-ray spectra. 

\begin{figure}[h]
  \centering
  \includegraphics[width=0.6\textwidth]{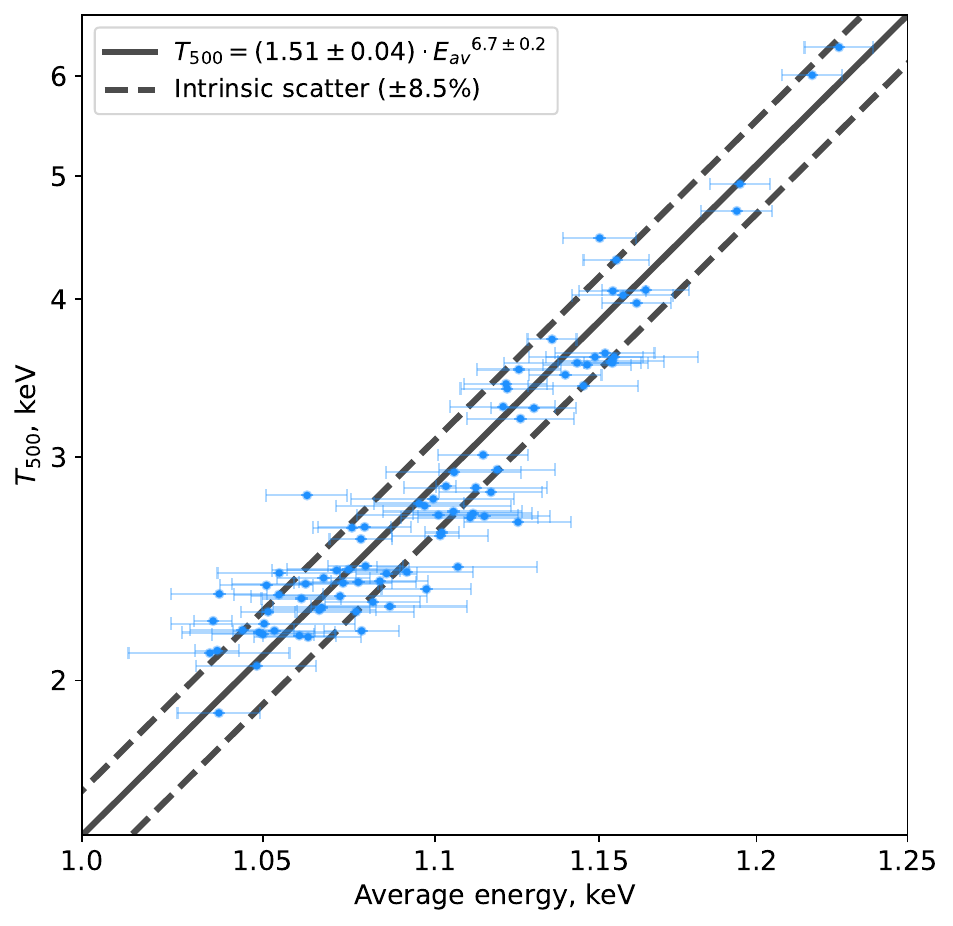}
  \caption{Temperature $T_{500}$ as a function of the average energy $E_{\rm av}$ in the $0.4-7.0$ keV band. Dashed lines mark the $68\%$ prediction band \hlc[yellow]{(intrinsic scatter)} for $T_{500}$. Note the absence of errors for the values of $T_{500}$ from the catalogue. The background is not taken into account.}
  \label{fig:T_vs_Eav}
\end{figure}

In Figure~\ref{fig:T_vs_Eav}, we compare the temperature $T_{500}$ and the average energy $E_{\rm av}$ (in the $0.4-7.0$ keV band) for each cluster in  the sample. \hlc[yellow]{There is a strong correlation between the temperature of a cluster and the average energy of its X-ray spectrum, which can be easily explained. For clusters with higher temperatures, the exponential cutoff in the spectrum shifts to higher energies. As a result, more X-ray photons are detected in the hard part of the spectrum, causing the average energy to increase.}
The temperature as a function of the average energy can be well described by a power law. The parameters values were obtained using the orthogonal distance regression (ODR); if not stated otherwise, hereinafter the same procedure is used for the power-law fitting. The best-fit function is also presented in Fig.~\ref{fig:T_vs_Eav} together with prediction bands for values of $T_{500}$. As a result, the value of the cluster temperature $T_{500}$ can be written as a function

\begin{equation}\label{eq:T_A.E.}
    T_{500} = (1.51 \pm 0.04) \ \text{keV} \Bigg( \frac{E_{\rm av}}{1 \text{keV}} \Bigg)^{6.7 \pm 0.2}
\end{equation}
with a scatter of $\sim 9\%$.

\begin{figure}[h]
  \centering
  \includegraphics[width=\textwidth]{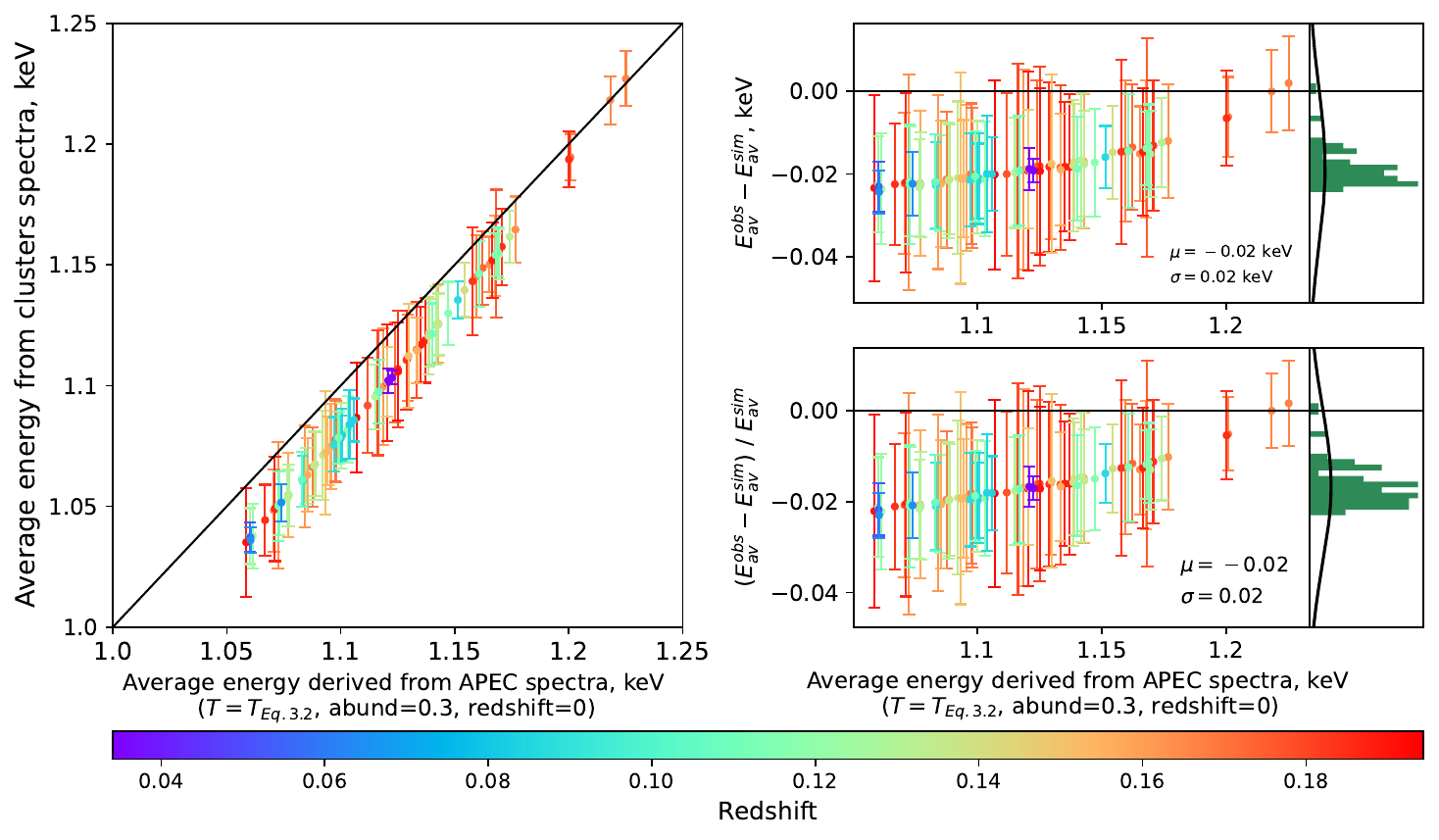}
  \caption{Comparison of average energies: $E_{\rm av}^{sim}$ from simulated single-temperature \texttt{APEC} models with a temperature equal to $T_{\rm spec}$ vs $E_{\rm av}^{obs}$ from spectral fitting for 84 clusters from \textit{Magneticum} simulations. The upper panel compares average energies directly, whereas the central panel compares absolute difference and the lower panel compares relative difference between them. The histograms on the right indicate mean values $\mu$ and scatter $\sigma$ for this comparisons. Redshifts are indicated by colorbar.}
  \label{fig:Eav_vs_Eav}
\end{figure}

\hlc[yellow]{However, the scatter of relationship }\ref{eq:T_A.E.}\hlc[yellow]{ would not be so tight if the spectrum were clearly multi-temperature.} To further illustrate the credibility of using the average energy as a temperature indicator, we \hlc[yellow]{perform a simple test.} Single-temperature spectra of each cluster are simulated using the \texttt{APEC} model with the following parameters: temperature is \hlc[yellow]{derived from Equation}~\ref{eq:T_A.E.} \hlc[yellow]{(based on average energy)}, abundance of heavy elements $Z=0.3$ of the Solar value {(the abundance tables from \cite{1989GeCoA..53..197A} were used)}, redshift $z=0$, \hlc[yellow]{normalization equal to 1; as previously, exposure time is set to $10 \ \rm ks$}. These spectra were then convolved with \textit{SRG/eROSITA} response matrices, and the average energy of the resulting spectrum in the $0.4-7.0$ keV band was calculated. In Figure~\ref{fig:Eav_vs_Eav}, these results are compared for each cluster with values of average energy computed from the cluster spectra extracted from the simulated photon lists. Both values are in good agreement, which indicates that the single-temperature model offers a very good approximation for the integral cluster spectra\hlc[yellow]{, especially after proper masking applied (see }Sec.~\ref{sec:im_gen}). \hlcolor[yellow]{However, one can see a small deviation (about 2\% difference in the values of the average energy) from the single-temperature model for observed clusters, since their spectra include emission from outer cooler regions. If we consider a smaller region, let's say a circle of radius $0.5 \cdot R_{500}$, then the cluster emission from this aperture is much better described by the single-temperature model.}

\hlcolor[yellow]{We recall that all the spectra of the observed clusters in our sample have been shifted to $z=0$. However, the average energy is a function of both temperature and redshift, and there is a certain degeneracy: the average energy is higher for higher temperatures or for lower redshifts.  Since in this work we consider only low-redshift clusters $z<0.2$, for such objects the dependence of the average energy on the redshift is rather weak, as shown in }Appendix~\ref{sec:aven-redshift}.

\subsection{Towards realistic spectra}\label{sec:background}

To simulate realistic observations of galaxy clusters with \textit{SRG/eROSITA} we add the astrophysical and particle X-ray background to previously obtained cluster spectra. For both components, we use empirical models that describe in-orbit background measurements \citep{2021A&A...647A...1P}. The details of the simulation of background are available in Appendix \ref{sec:background_model}. To accurately estimate the contribution of different components to the total observed spectra, we assume that the count rates from both the astrophysical and particle background are known with high certainty. For this reason, we simulate mock background observations with sufficiently long exposure of $10^4$ ks (for one telescope module), which then can be properly added to each cluster (i.e. considering its angular size).

\begin{figure}[h]
  \centering
  \includegraphics[width=\textwidth]{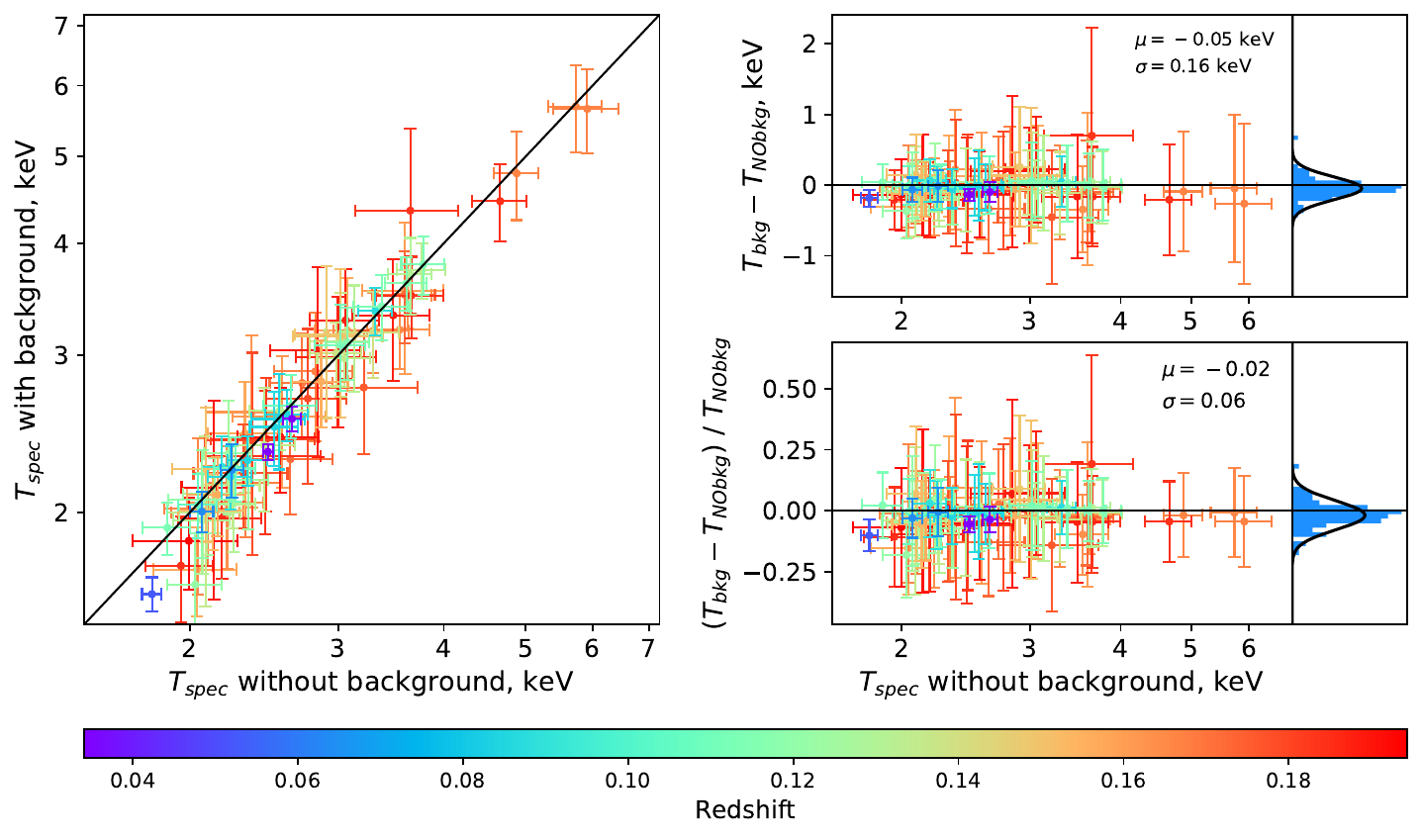}
  \caption{ Comparison of temperatures: $T_{\rm spec}$ (without background) vs. $T_{\rm spec}^{BKG}$ (with background). The upper panel compares temperatures directly, whereas central panel compares absolute difference and lower panel compares relative difference between them. The histograms on the right indicate mean values $\mu$ and scatter $\sigma$ for this comparisons. Redshifts are indicated by colorbar. Note the log-scale for horizontal axis.}
  \label{fig:T_with_bkg_vs_T_without_bkg}
\end{figure}

For each cluster, the total spectra (cluster + background) is simulated and fitted the same way as described in Section~\ref{sec:overview}, and its average energy $E_{\rm av}^{tot}$ (again, in $0.4-7.0$ keV band) is calculated. After averaging values of spectral temperatures with background added (we denote this value as $T_{\rm spec}^{BKG}$) the results are compared with previously obtained values of $T_{\rm spec}$ without background. Temperature comparison is shown in Figure~\ref{fig:T_with_bkg_vs_T_without_bkg}. The differences between values of spectral temperature with and without background are very small (this is a consequence of our confidence in the accuracy of the background model), but, however, uncertainties are noticeably larger than for the case without background.

Given the expected count rates from both astrophysical and particle background, we can estimate their fractions in total count rate, which are denoted as $n_{\rm ph}$ and $n_{\rm pbkg}$ respectively, as well as their average energies $E_{\rm av}^{ph}$ and $E_{\rm av}^{pbkg}$. In this case, the expression for the average energy of cluster spectra is  

\begin{equation} \label{eq:aven_bkg}
    E_{\rm av}^{*} = \frac{E_{\rm av}^{tot} - E_{\rm av}^{ph} \cdot n_{\rm ph} - E_{\rm av}^{pbkg} \cdot n_{\rm pbkg}}{1 - n_{\rm ph} - n_{\rm pbkg}}
\end{equation}

\begin{figure}[h]
  \centering
  \includegraphics[width=0.6\textwidth]{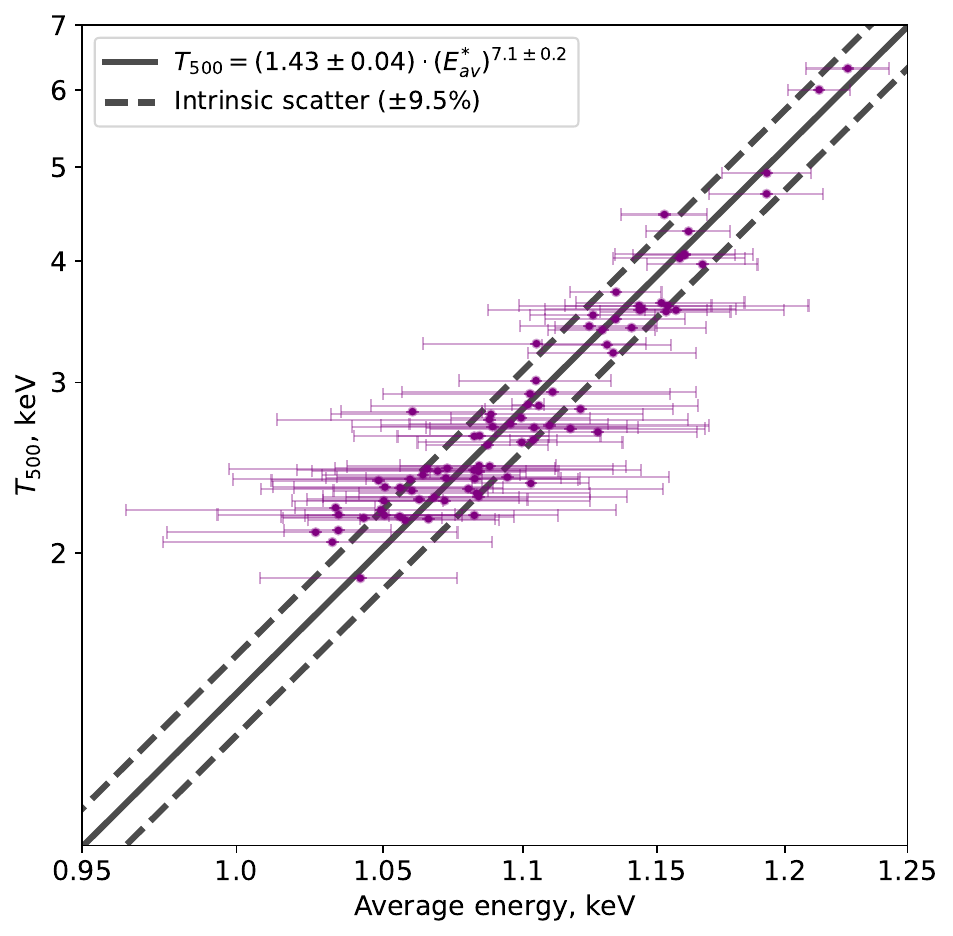}
  \caption{Temperature $T_{500}$ as function of average energy $E_{\rm av}^{*}$ in the $0.4-7.0$ keV band. Dashed lines show $68\%$ prediction bands \hlc[yellow]{(intrinsic scatter)} for $T_{500}$. The astrophysical and instrument backgrounds are taken into account in this case.}
  \label{fig:T_vs_Eav_with_bkg}
\end{figure}

The comparison of $T_{500}$ from the catalogue with average energy $E_{\rm av}^{*}$ of a cluster in the presence of background and derivation of expression for it is performed similar to Section~\ref{sec:average_energy} and is shown in Figure~\ref{fig:T_vs_Eav_with_bkg}. As earlier, the dependence of temperature $T_{500}$ as a function of average energy is well described by power-law

\begin{equation}\label{eq:T_A.E._BKG}
    T_{500}^{BKG} = (1.43 \pm 0.04) \ \text{keV} \Bigg( \frac{E_{\rm av}^{*}}{1~\text{keV}} \Bigg)^{7.1 \pm 0.2}
\end{equation}

with slightly larger scatter than in the case of no background (but still of $\sim10\%$ order). The best-fit function is also presented in Fig.~\ref{fig:T_vs_Eav_with_bkg} together with the confidence and prediction bands. This function is in good agreement with Equation~\ref{eq:T_A.E.}, even with visibly larger errors in $E_{\rm av}$ for the fixed value of temperature. Thus, given a proper accounting for the background, the average energy shows robustness as an indicator of a temperature.

\section{Cluster mass estimations}\label{sec:discussion}

\begin{figure}
\centering
\includegraphics[height=0.44\textwidth, trim = 0cm 0cm 1cm 2cm, clip]{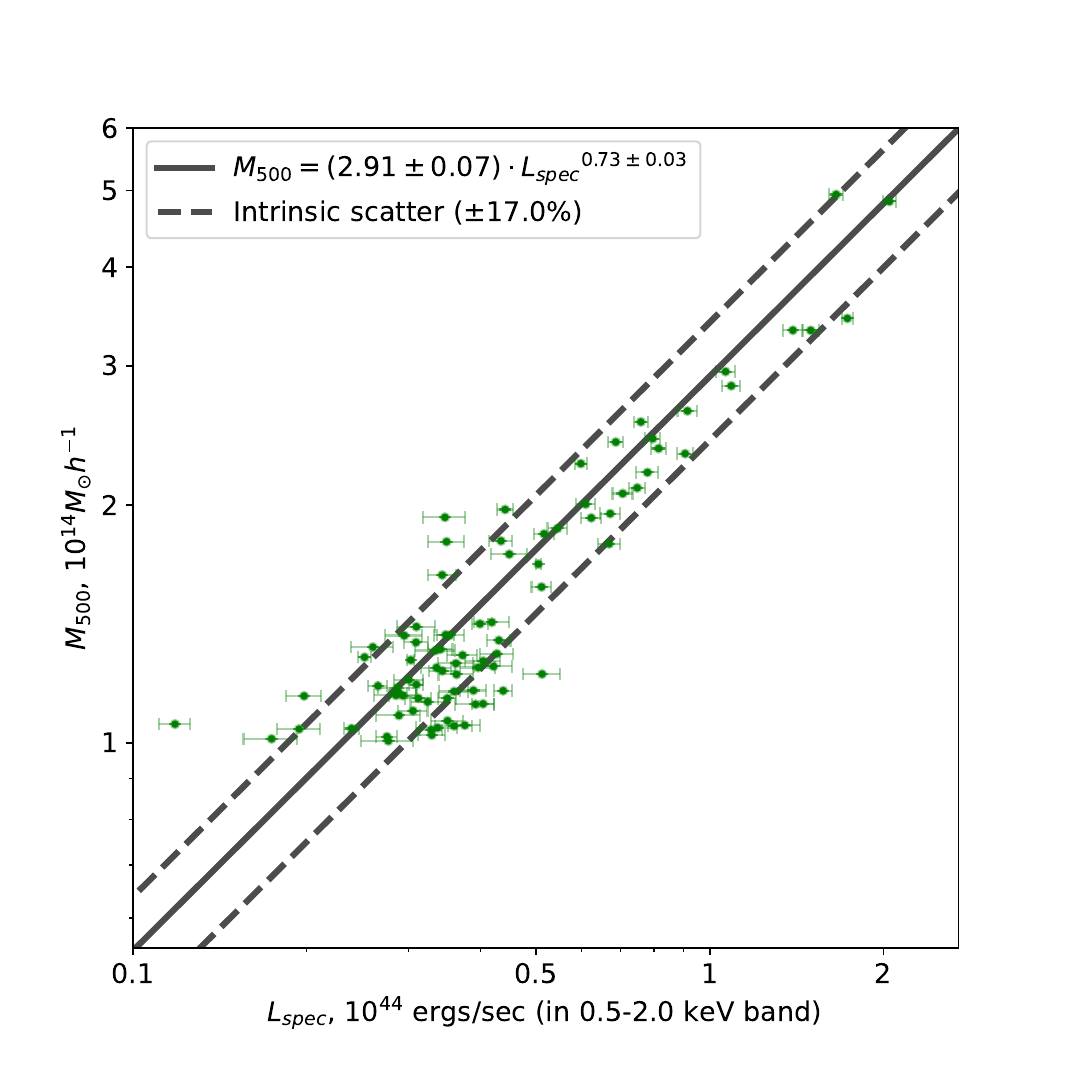} \\
\includegraphics[height=0.44\textwidth, trim = 0cm 0cm 1cm 2cm, clip]{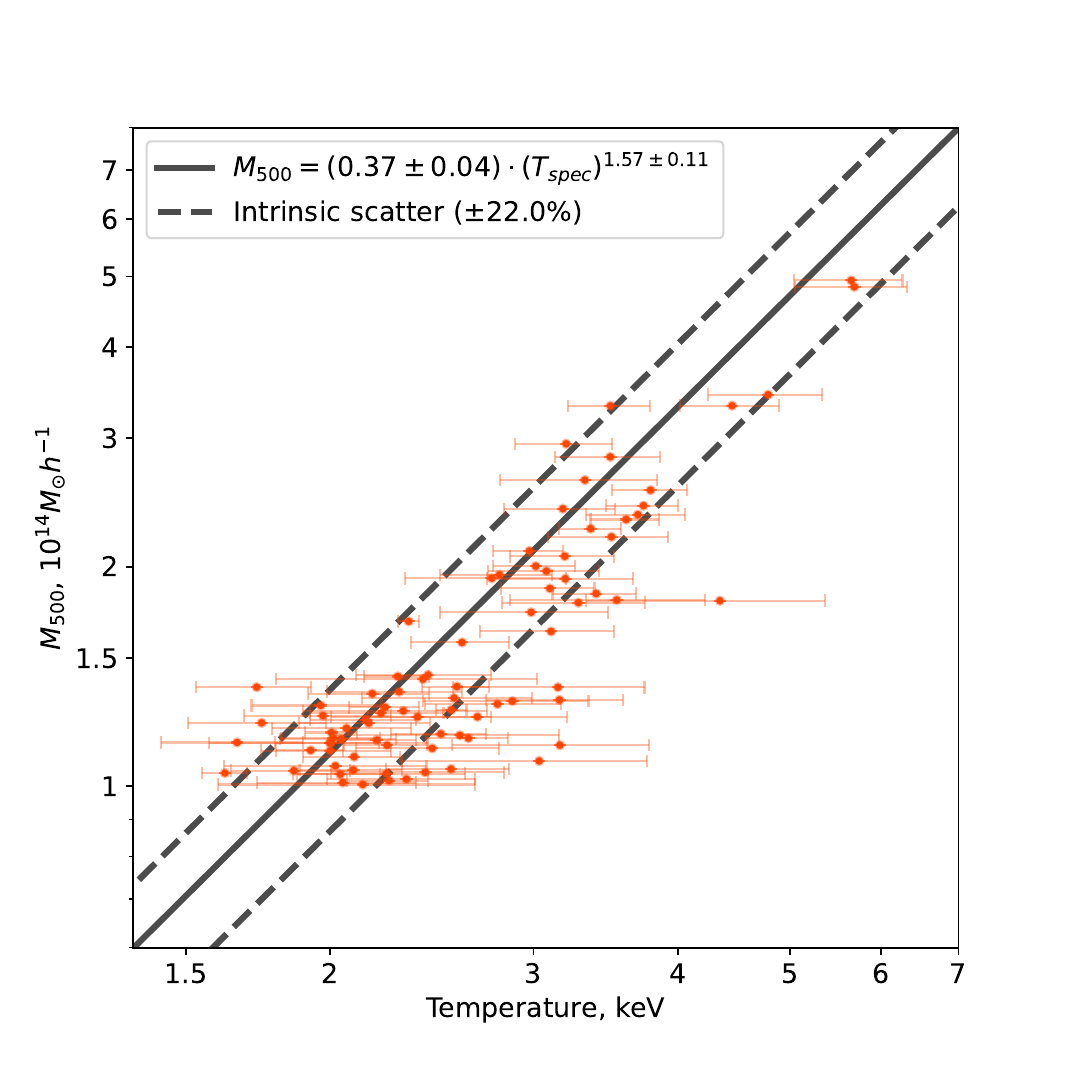} \\
\includegraphics[height=0.44\textwidth, trim = 0cm 0cm 1cm 2cm, clip]{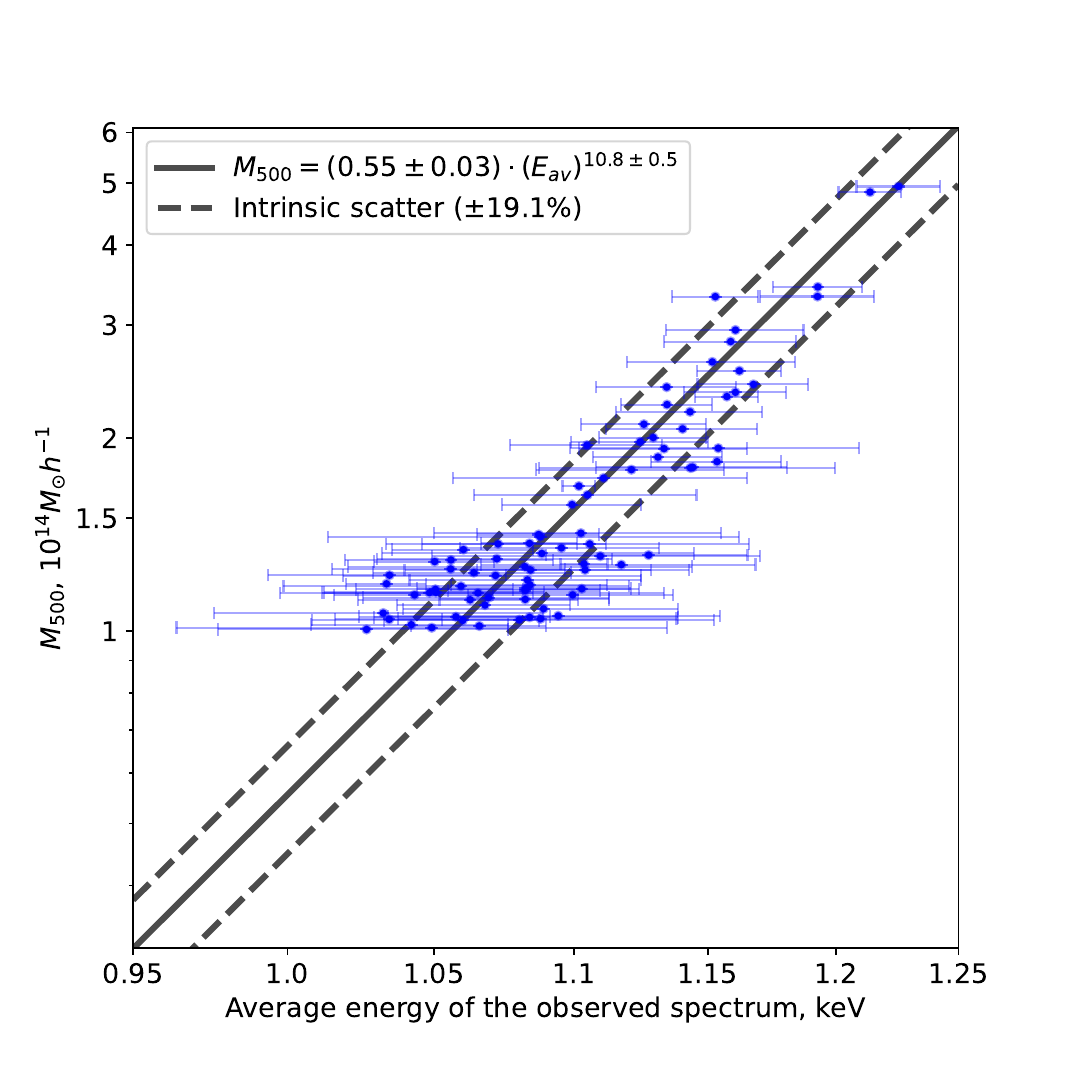}
\caption{\textit{From top to bottom:} $M_{500}$ as a function of luminosity $L_{\rm spec}$ (in $0.5-2.0$ keV band), spectral temperature $T_{\rm spec}$ and average energy $E_{\rm av}$ (in $0.4-7.0$ keV band). Dashed lines show $68\%$ prediction band.}
\label{fig:results}
\end{figure}

To frame the cosmological implications of our work, we compare the scatter of different proxies for mass with each other. At first, we recall the calculation of the X-ray luminosity $L_{\rm spec}$ in the $0.5-2.0$ keV band, which was performed simultaneously with spectral fitting (it was described in Section~\ref{sec:overview}). In Figure~\ref{fig:results} (upper panel) we show the $M_{500}-L_{\rm spec}$ relation for our cluster sample. This could be well described by the expression

\begin{equation}\label{eq:M_vs_L}
     M_{500} = (2.91 \pm 0.07) \cdot 10^{14} M_{\odot} h^{-1} \ \Bigg( \frac{L_{\rm spec}}{10^{44} \ \text{ergs/s}} \Bigg)^{0.73 \pm 0.03}
\end{equation}

with the scatter of $17\%$ for a fixed value of luminosity. The scatter of this expression, if reversed, translates into $\sim 24\%$ spread of luminosity for a fixed value of the mass, which is a sufficiently large but expected value. \hlc[yellow]{This dependence shows less scatter than the one implied by Equation}~\ref{eq:L_X} \hlc[yellow]{and has a somewhat smaller slope.}

Next, we demonstrate the $M_{500}-T_{\rm spec}$ relation, which is shown in Figure~\ref{fig:results} (central panel) and is given by the expression

\begin{equation}\label{eq:M_vs_T}
     M_{500} = (0.37 \pm 0.04) \cdot 10^{14} M_{\odot} h^{-1} \ \Bigg( \frac{T_{\rm spec}}{1 \ \text{keV}} \Bigg)^{1.57 \pm 0.11}
\end{equation}

with $22\%$ scatter. \hlc[yellow]{Being inverted, this expression agrees well with} Equation~\ref{eq:T_X}, \hlc[yellow]{which was previously shown in} Figure~\ref{fig:scaling_relations} \hlc[yellow]{for $M_{500}-T_{500}$ relation.}

Finally, we present the $M_{500}-E_{\rm av}$ relation. It is also shown in Figure~\ref{fig:results} (bottom panel) and is described by the expression

\begin{equation}\label{eq:M_vs_Eav}
    M_{500} = (0.55 \pm 0.03) \cdot 10^{14} M_{\odot} h^{-1} \Bigg( \frac{E_{\rm av}}{1 \text{keV}} \Bigg)^{10.8 \pm 0.5}
\end{equation}

with the scatter of $\sim 19\%$. \hlc[yellow]{As for the $M_{500}-T_{\rm spec}$ and $M_{500}-L_{\rm spec}$ relations, there is hardly any difference between the use of the $M_{500}-E_{\rm av}$ relation obtained from the spectra with or without background.} Having similar scatter, $M_{500}-E_{\rm av}$ relation is much easier to obtain because it does not require spectral fitting.

It is a question of interest whether the obtained relations remain the same in case of low count statistics. We have performed the same procedure as before, but the exposure time was reduced to 1 ks of observations for each cluster. As the statistics deteriorate, the measurement scatter of $L_{\rm spec}, T_{\rm spec}$ and $E_{\rm av}$ for a fixed mass value grows, but, surprisingly, the scaling relations and their intrinsic scatter keep almost the same.

Another problem that needs further investigation is selection effects. It can be clearly seen that in each panel of Figure~\ref{fig:results} the cluster sample can be divided into two sub-samples. The low-mass group ($M_{500}\simeq(1-1.5)\cdot 10^{14} M_{\odot} h^{-1}$) tends to shift the overall dependence to higher values of the observable quantities for a fixed mass, whereas the high mass group ($M_{500} > 1.5\cdot 10^{14} M_{\odot} h^{-1}$) obeys the general slope quite well.

\section{Conclusions}\label{sec:conclusions}
 
Using a sample of 84 massive galaxy clusters from the \textit{Magneticum} cosmological hydrodynamical simulations, we analysed spectra of their predicted X-ray emission integrated over $R_{500}$ radii and mocked into observational data corresponding to the response functions and exposure time of the \textit{SRG/eROSITA} all-sky survey (including astrophysical and instrumental backgrounds). A simple image-based filtering of the registered counts allows us to obtain an almost uncontaminated ICM spectrum (at the affordable cost of losing $\lesssim25\%$ of the counts), which is shown to be quite well described by a single-temperature thermal emission model.

The average energy $E_{\rm av}$ of the detected X-ray spectrum of a cluster in the $0.4-7.0$ keV band turns out to be a stable proxy for the ICM mass-weighted temperature $T_{500}$, being model-independent and insensitive to the properly accounted background contribution.

The relations between the cluster mass $M_{500}$, its luminosity $L_{\rm spec}$ (in the $0.5-2.0$ keV band), spectral temperature $T_{\rm spec}$ and average energy $E_{\rm av}$ (in the $0.4-7.0$ keV band) have less than $20\%$ statistical scatter, allowing one to use $E_{\rm av}$ as a reliable and simple (not requiring spectral fitting) observational proxy. Although the exact form of the derived relations obviously depends on the parameters of the X-ray telescope, they can be readily calibrated using the same procedure and taking advantage of the clusters' spectral database, which is made publicly available with the current study.

\appendix

\section{Database of X-ray images and spectra of clusters}\label{sec:database}

\begin{figure*}
  \hspace*{+0.3cm}\includegraphics[width=0.98\textwidth]{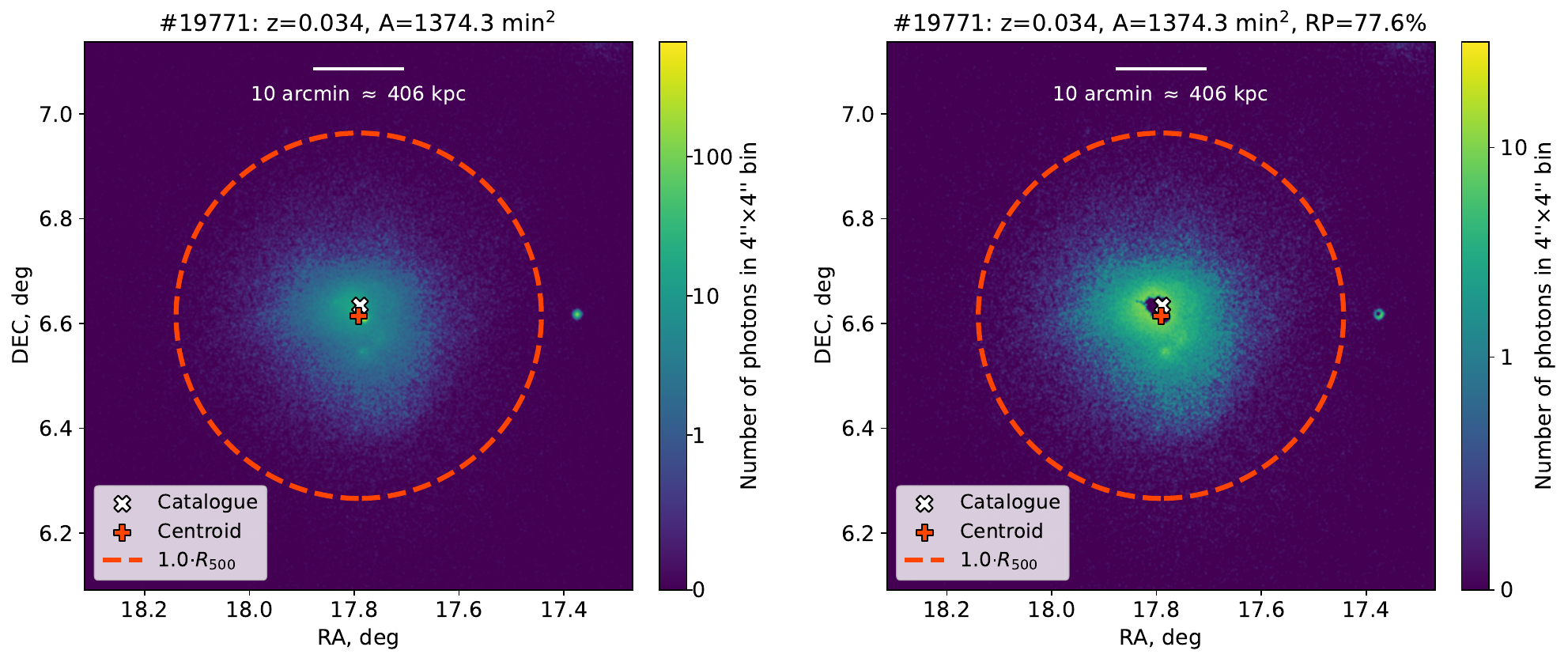}
  \includegraphics[width=\textwidth]{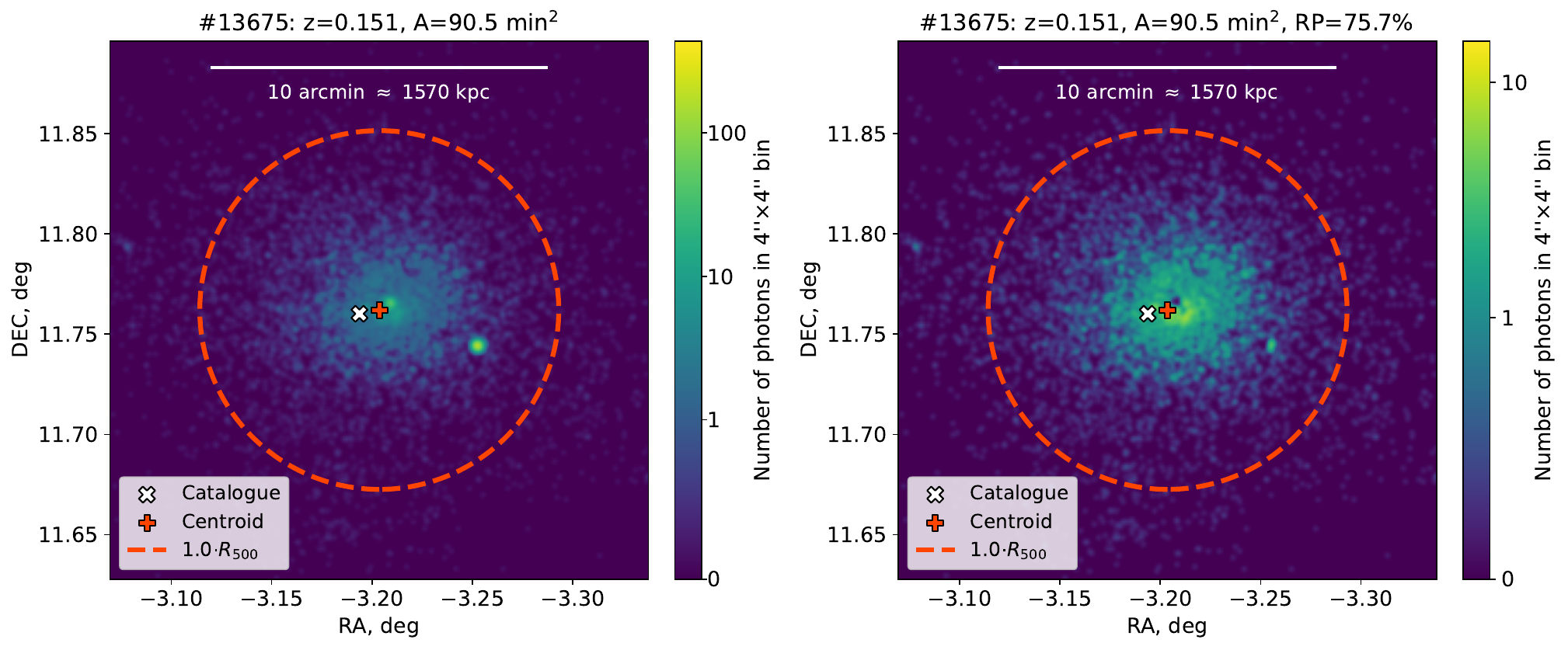}
  \includegraphics[width=\textwidth]{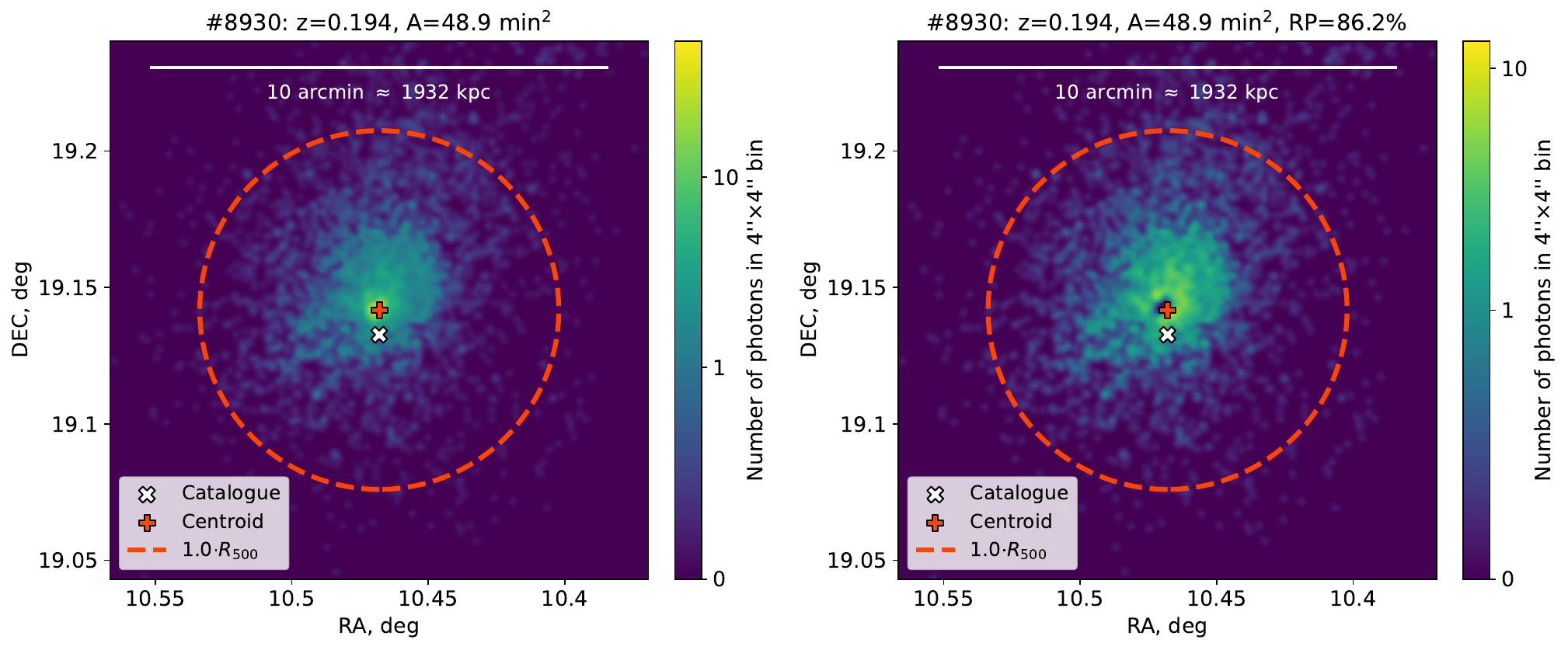}
  \caption{\hlcolor[yellow]{Examples of clusters images at $z_{min}=0.034, z_{median}=0.151 \ \text{and} \ z_{max}=0.194$ (from top to bottom).} The left panel in each row shows image obtained with all the photons from the corresponding photon list, whereas the right panels show filtered images. $A$ is the area of circle with radius equal to $R_{500}$ and $RP$ is the fraction of remaining photons after filtering.}
  \label{fig:images}
\end{figure*}

Examples of original and filtered images of three different studied galaxy clusters at $z = \{0.035, 0.089, 0.157\}$ are shown in Figure~\ref{fig:images}. The filtering procedure is described in Section~\ref{sec:im_gen}. In each image, there is a circle with a radius equal to $R_{500}$ for this cluster with its center located at the centroid of the image; there is also a point indicating the position of the gravitational potential minimum from \textit{Magneticum} catalogue.

We present the database which contains all 84 clusters' images (both unfiltered and filtered) and their spectra. The description of the image development and filtering procedure is given in Section~\ref{sec:im_gen}. After filtering, for each cluster the extracted photons are binned by energies over a range from $0.1$ to $12.0$ keV in $4096$ logarithmic steps, resulting in flux density spectra. All model spectra are stored in a single FITS file, which can be passed as an input to \texttt{XSPEC} software \citep{1996ASPC..101...17A}. For spectral models in this database, the overall normalization has been set as if each cluster were $100$ Mpc distant.

\begin{figure*}[h]
  \includegraphics[width=\textwidth]{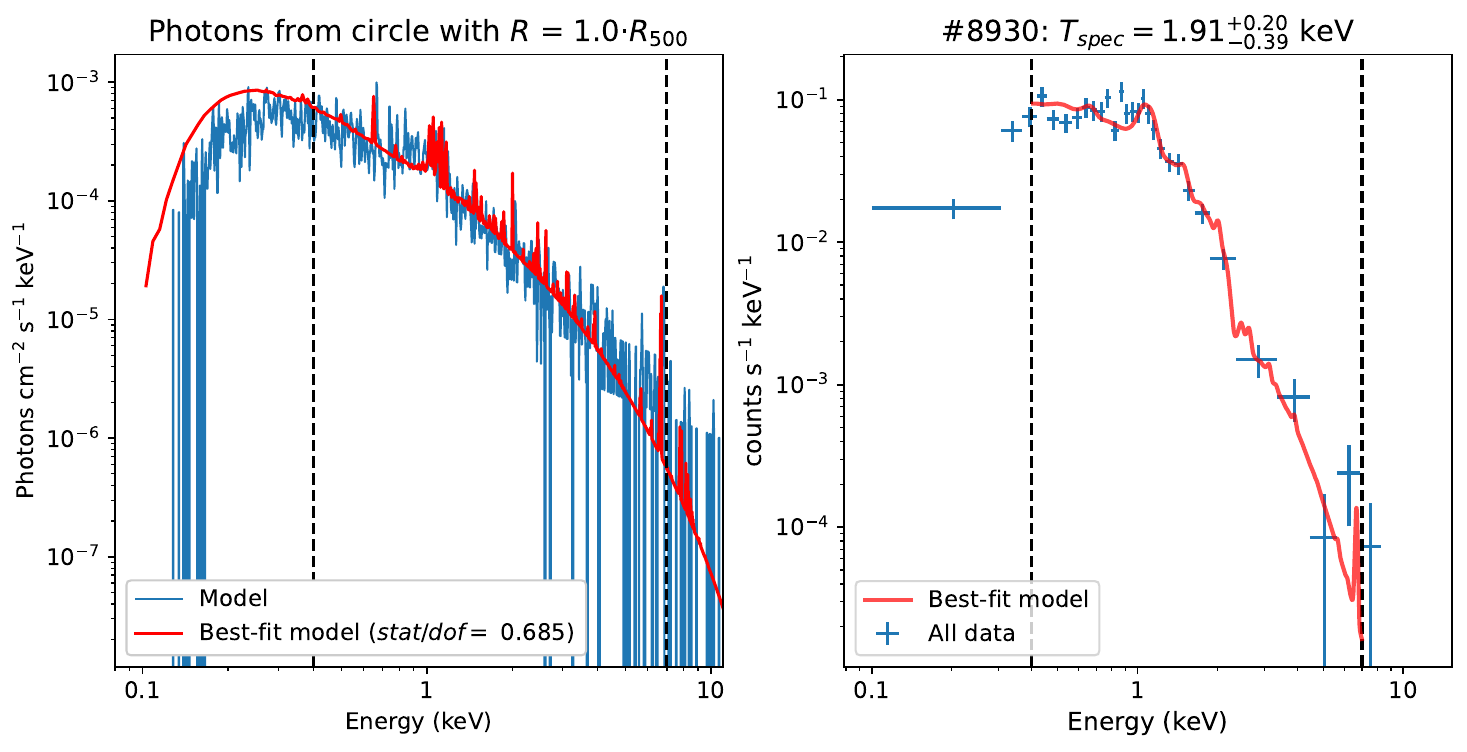}
  \includegraphics[width=\textwidth]{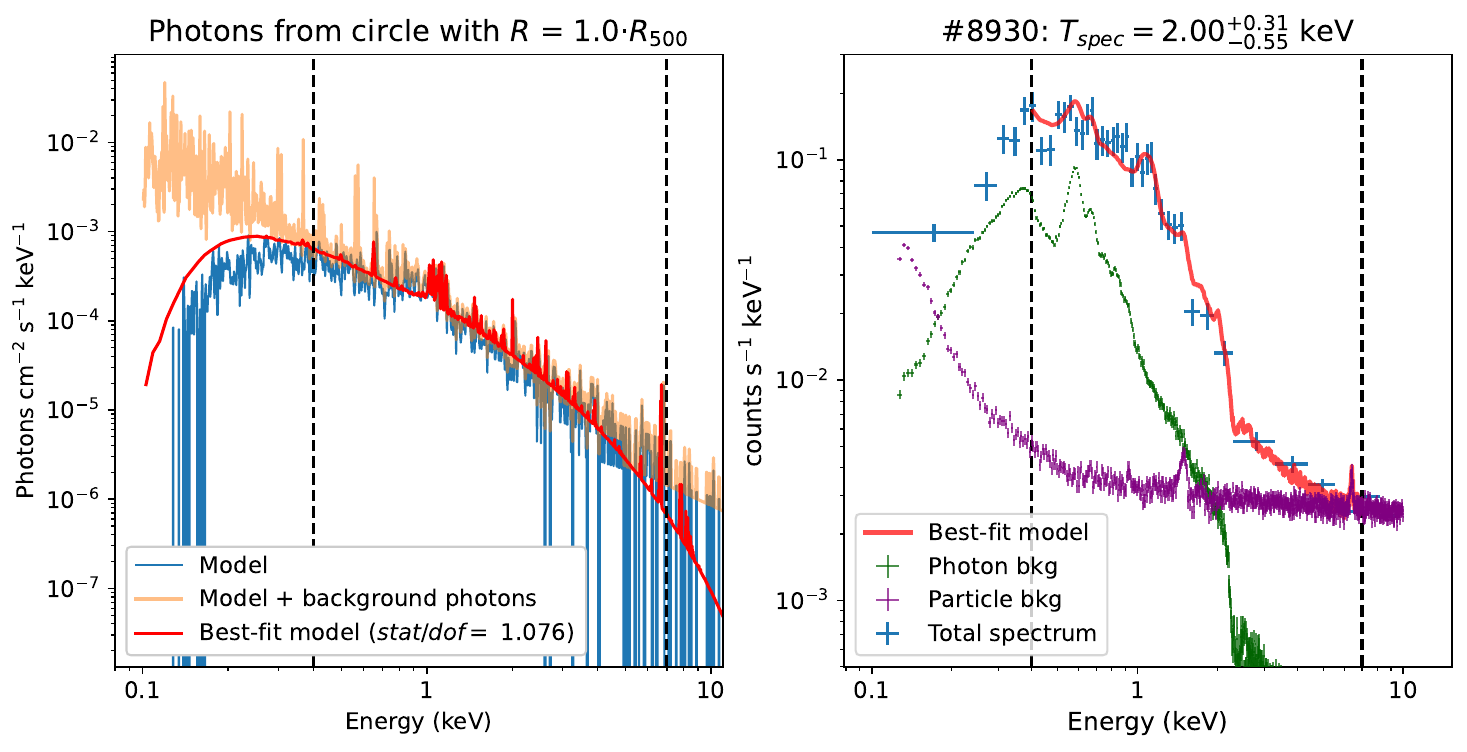}
  \caption{Examples of spectral analysis of cluster \hlcolor[yellow]{$\#8930$ ($z_{true}=0.194$, $M_{500} = 1.05 \cdot 10^{14} M_{\odot}/h$, $T_{500}=2.29$ keV)}. Top and bottom rows show cases without and with background, respectively. \hlc[yellow]{In each row, the left panel shows the model spectra (which is not convolved with any instrument) and the right panel shows \textit{SRG/eROSITA} mock spectrum. Both panels also show the best-fit model obtained by fitting the observed spectrum in the $0.4-7.0$ keV energy band.} The right panel of the bottom row also shows the contributions of \hlc[yellow]{astrophysical and particle background components in the total spectrum (their normalisation is determined by the area of the cluster in the sky)}.}
  \label{fig:spectra}
\end{figure*}

Examples of X-ray cluster spectra along with the best-fit model and added background are shown in Figure~\ref{fig:spectra}. The fitting procedure is described in Section~\ref{sec:overview}.

\section{Estimation of the statistical uncertainties}\label{sec:howmuch}

\begin{figure}[h]
  \centering
  \includegraphics[height=0.96\textwidth]{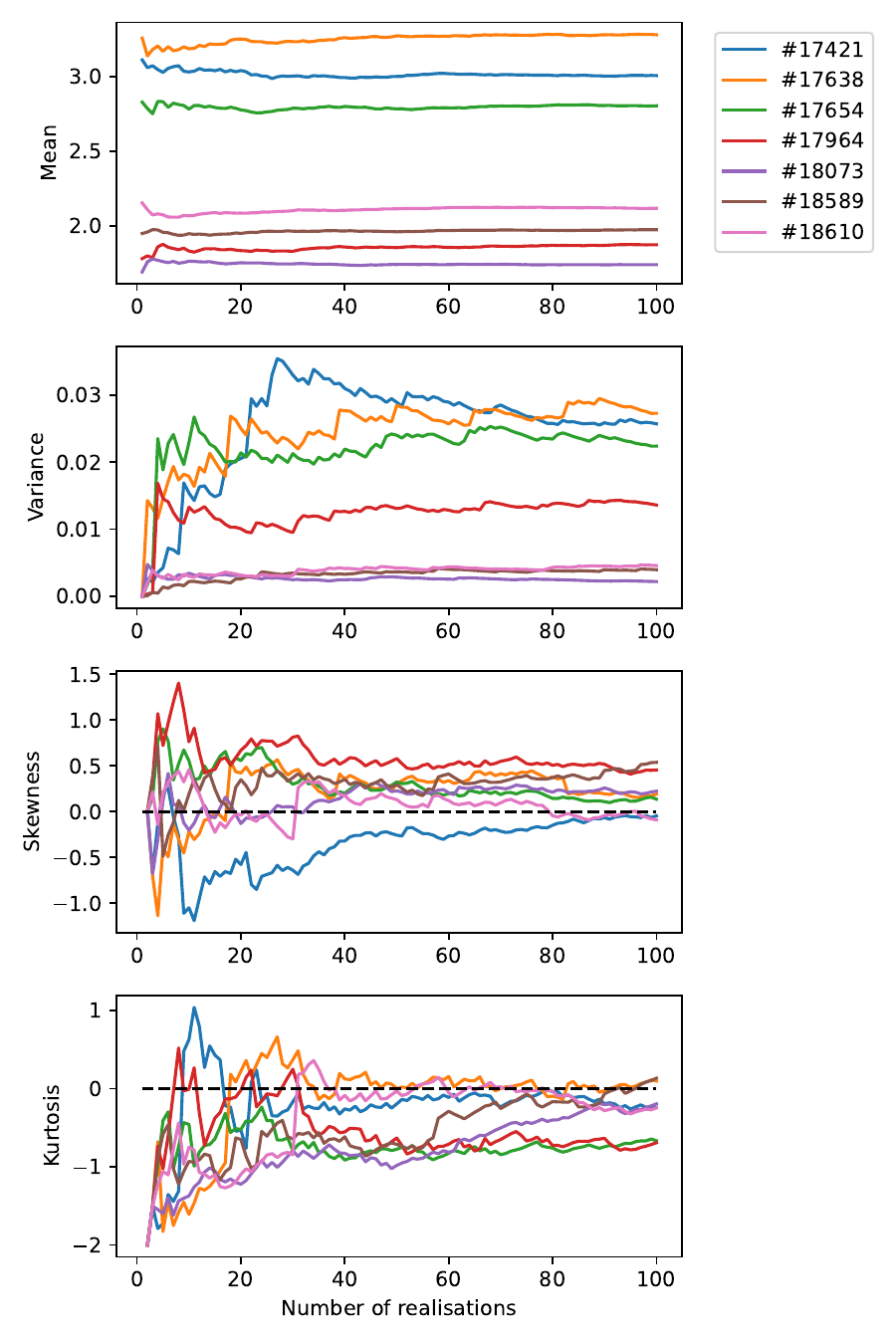}
  \caption{Moments of temperature distribution from first (\textit{upper panel}) to fourth (\textit{bottom panel}) as a functions of increasing number of fitting realisations. Different colors indicate 7 clusters and correspond to their numbers in \textit{Magneticum} catalogue.}
  \label{fig:moments}
\end{figure}

To reduce statistical uncertainties in values of temperature, luminosity and average energy, simulated spectrum fitting procedures are performed 50 times for each cluster, and the mean values of spectral temperature distribution were considered as true values. Concurrently, we calculated the second, third and fourth moments of these distributions (variance, skewness and kurtosis, respectively). In Figure~\ref{fig:moments} we present values of these moments as functions of an increasing number of fitting realisations (up to 100) for 7 distinctive clusters. While the mean value is reached rather quickly as averaging proceeds, other moments demonstrate non-normal distribution behaviour even in case of sufficiently enough realisations. The same procedure was performed simultaneously for luminosity and average energy, and they demonstrate similar behaviour.

\section{Models for background}\label{sec:background_model}

To simulate astrophysical and particle X-ray background, we have used empirical models that describe in-orbit background measurements performed by \textit{SRG/eROSITA} \citep{2021A&A...647A...1P}.

In particular, the model for the particle background we implement is derived from the released filter wheel closed data measurements \citep{2021A&A...647A...1P}, which have been shown to provide excellent agreement with the survey-averaged instrumental background level \citep[e.g.][]{2023MNRAS.525..898L}.

The model for the astrophysical background consists of four components, including Local Bubble emission, Galactic hot halo and ''corona'' X-ray Galactic emission, and unresolved point X-ray sources, i.e. Cosmic X-ray background (CXB, we assume that only half of it is resolved), as motivated by previous observations by \textit{Chandra} \citep{2006ApJ...645...95H, 2007ApJ...661L.117H, 2007ApJ...671.1523H, 2003ApJ...583...70M} and \textit{SRG/eROSITA} \citep{2023A&A...674A.195P}. The first three components can be well described by a thin-thermal plasma and modeled by \texttt{APEC} model with temperatures and normalizations equal to (0.099 keV, 0.00174), (0.225 keV, 0.0007475), (0.7 keV, 0.0000897), redshift equal to 0 and metallicity equal to 1 Solar, Galactic absorbing column equal to $0.018 \cdot 10^{22} \ \rm cm^{-2}$. The latter component is described by a power law with a photon index equal to 1.47 and normalization equal to 0.0001052. We have implemented the photon background as the model for \texttt{XSPEC} in the following form:
\begin{multline}
    \texttt{const(1)*(apec(2)+ wabs(3)*apec(4)+apec(5)+const(6)*powerlaw(7)))}
\end{multline}

This model is normalized for 1 min$^2$. To simulate the background flux corresponding to the aperture of a particular cluster, the flux from this model must be multiplied by the projected area of its radius $R_{500}$ in square minutes. During the spectral fitting procedure, all components are frozen, but overall normalisation is considered as a free parameter.

\section{Average energy as a function of redshift}\label{sec:aven-redshift}

\begin{figure}[h]
  \centering
  \includegraphics[height=0.5\textwidth]{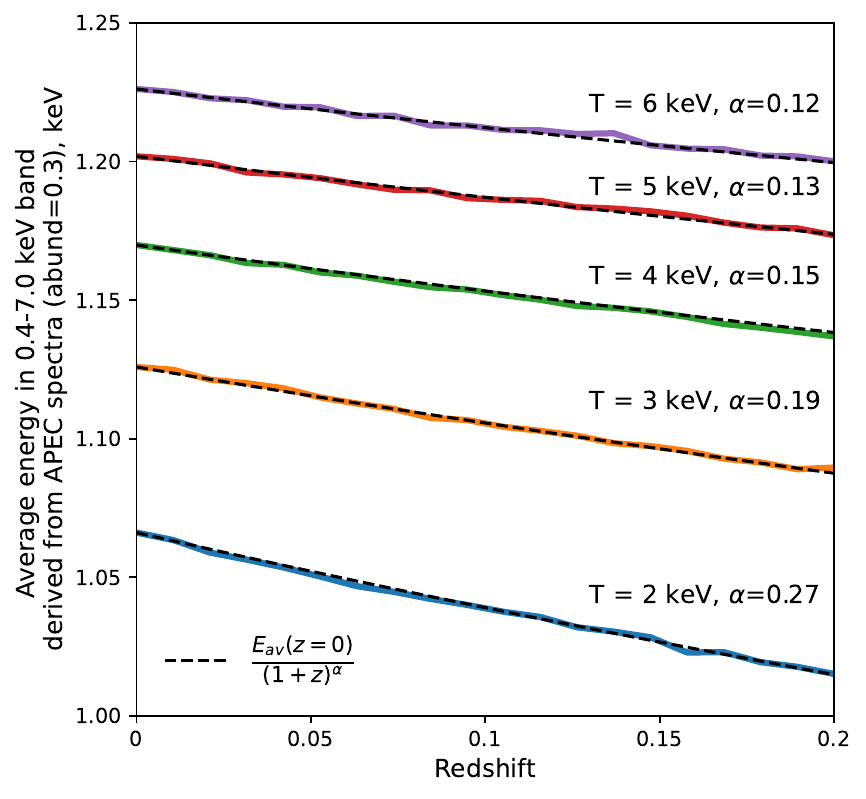}
  \caption{\hlcolor[yellow]{Average energy as a function of redshift for different temperatures corresponding to the sample from the \textit{Magneticum} lightcone. The solid lines show the dependence for simulated \texttt{APEC} spectra convolved with \textit{SRG/eROSITA} response matrices, while the dashed lines show the analytical approximations.}}
  \label{fig:avenz}
\end{figure}

\hlcolor[yellow]{To demonstrate the effect of the unknown redshift, we have simulated the \texttt{APEC} spectra convolved with \textit{SRG/eROSITA} response matrices for a set of temperatures and redshifts corresponding to the clusters in our sample. The dependence of the average energy $E_{av}$ on the redshift is shown in} Fig.~\ref{fig:avenz}. \hlcolor[yellow]{Since $E_{av}$ as a function of $z$ is proportional to $1/(1+z)^\alpha$, where $\alpha=[0.12, 0.27]$, and $T \sim E_{av}^{6.7}$, the uncertainty in the derived values of $T$ will be proportional to $\sim 1/(1+z)$ and comparable to the scatter for} Eq.~\ref{eq:T_A.E.}.

\acknowledgments

We are thankful to Eugene Churazov, Alexey Vikhlinin and Ilaria Marini for valuable discussions and suggestions. \hlc[yellow]{We thank the anonymous referee for helpful comments that improved the quality of this work.}

IK and KD acknowledge support by the COMPLEX project from the European Research Council (ERC) under the European Union’s Horizon 2020 research and innovation program grant agreement ERC-2019-AdG 882679.

The catalogue of \textit{Magneticum} clusters and the corresponding photon lists are publicly available online. The extracted ({un}filtered and filtered for bright substructures) cluster images and Radially-Resolved Clusters Spectral Database are available online\footnote{ \href{https://github.com/pi4imu/RRCS_DB}{\texttt{https://github.com/pi4imu/RRCS\_DB}}}.



\bibliographystyle{JHEP}
\bibliography{biblio.bib}







\end{document}